\documentclass[fleqn,usenatbib]{mnras}

\usepackage{newtxtext,newtxmath}

\usepackage[T1]{fontenc}

\DeclareRobustCommand{\VAN}[3]{\#2}
\let\VANthebibliography\thebibliography
\def\thebibliography{\DeclareRobustCommand{\VAN}[3]{\#\#3}\VANthebibliography}

\usepackage{graphicx}	
\usepackage{amsmath}	
\usepackage{glossaries}
\usepackage{longtable}
\usepackage{pdflscape}
\usepackage{rotating}

\usepackage{xspace}

\renewcommand{\eqref}[1]{Eq.\@~\ref{\#1}\xspace}
\newcommand{\kepler}{\emph{Kepler}\xspace}
\newcommand{\TESS}{\emph{TESS}\xspace}
\newcommand{\kt}{\emph{K2}\xspace}

\newcommand{\Gaia}{\emph{Gaia}\xspace}
\newcommand{\TMASS}{\textsc{2MASS}\xspace}
\newcommand{\BASTA}{\textsc{basta}\xspace}
\newcommand{\APOGEE}{\textsc{APOGEE}\xspace}

\newcommand{\Garstec}{\textsc{Garstec}\xspace}
\newcommand{\PBJam}{\textsc{PBjam}\xspace}
\newcommand{\SYD}{\textsc{syd}\xspace}

\newcommand{\numax}{$\nu_{\mathrm{max}}$\xspace}
\newcommand{\dnu}{$\Delta \nu$\xspace}
\newcommand{\alphafe}{[$\alpha$/Fe]\xspace}
\newcommand{\teff}{$T_{\mathrm{eff}}$\xspace}

\newcommand{\En}{$E$\xspace}
\newcommand{\Lz}{$L_z$\xspace}

\newcommand{\nuindi}{$\nu$ Indi\xspace}
\newcommand{\Enc}{GES\xspace}
\newcommand{\finalage}{$9.5_{-1.3}^{+1.2}$ Gyr\xspace}
\newcommand{\Montalban}{M21\xspace}

\newcommand{\RG}{red giant\xspace}

\title[Age Determination of Merger Remnant Stars]{Age Determination of Galaxy Merger Remnant Stars using Asteroseismology}

\author[Camilla~C.~Borre et al.]{Camilla~C.~Borre,$^{1}$\thanks{E-mail: cborre@phys.au.dk}
V\'{i}ctor Aguirre Børsen-Koch,$^{1}$
Amina Helmi,$^{2}$ 
Helmer~H.~Koppelman,$^{3}$ \newauthor
Martin~B.~Nielsen,$^{4}$ 
Jakob~L.~Rørsted,$^{1}$
Dennis Stello,$^{1,5,6,7}$ 
Amalie Stokholm,$^{8,1}$
Mark L. Winther,$^{1}$\newauthor
Guy R. Davies,$^{4}$
Marc Hon,$^{9}$
J.~M.~Diederik Kruijssen,$^{10}$
Chervin Laporte,$^{11}$ 
Claudia Reyes,$^{5}$ 
Jie Yu,$^{12}$
\\
$^{1}$Stellar Astrophysics Centre, Department of Physics and Astronomy, Aarhus University, Ny Munkegade 120, DK-8000 Aarhus C, Denmark\\
$^{2}$Kapteyn Astronomical Institute, University of Groningen, Landleven 12, 9747 AD Groningen, The Netherlands\\
$^{3}$School of Natural Sciences, Institute for Advanced Study, 1 Einstein Drive, Princeton, NJ 08540, USA\\
$^{4}$School of Physics and Astronomy, University of Birmingham, Birmingham B15 2TT, UK\\
$^{5}$School of Physics, The University of New South Wales, Sydney NSW 2052, Australia\\
$^{6}$ARC Centre of Excellence for Astrophysics in Three Dimensions (ASTRO-3D), Australia\\
$^{7}$Sydney Institute for Astronomy (SIfA), School of Physics, University of Sydney, NSW 2006, Australia\\
$^{8}$Dipartimento di Fisica e Astronomia, Universit\`a degli Studi di Bologna, Via Gobetti 93/2, I-40129 Bologna, Italy\\
$^{9}$Institute for Astronomy, University of Hawai`i,2680 Woodlawn Drive, Honolulu, HI 96822, USA\\
$^{10}$Astronomisches Rechen-Institut, Zentrum f\" ur Astronomie der Universit\"at Heidelberg, M\"onchhofstra\ss e 12-14, D-69120 Heidelberg, Germany \\
$^{11}$Institut de Ciencies del Cosmos (ICCUB), Universitat de Barcelona (IEEC-UB), Mart\'{i}ı i Franques 1, E-08028 Barcelona, Spain\\
$^{12}$Max-Planck-Institut für Sonnensystemforschung, Justus-von-Liebig-Weg 3, 37077, Göttingen, Germany 
}

\date{Accepted XXX. Received YYY; in original form ZZZ}

\pubyear{2021}

\begin{document}
\label{firstpage}
\pagerange{\pageref{firstpage}--\pageref{lastpage}}
\maketitle 

\begin{abstract}
The Milky Way was shaped by the mergers with several galaxies in the past. We search for remnant stars that were born in these foreign galaxies and assess their ages in an effort to put upper limits on the merger times and thereby better understand the evolutionary history of our Galaxy. 
Using 6D-phase space information from \Gaia eDR3 and chemical information from \APOGEE DR16, we kinematically and chemically select $23$ red giant stars belonging to former dwarf galaxies that merged with the Milky Way. With added asteroseismology from \kepler and \kt, we determine the ages of the $23$ ex-situ stars and $55$ in-situ stars with great precision. 
We find that all the ex-situ stars are consistent with being older than 8 Gyr. While it is not possible to associate all the stars with a specific dwarf galaxy we classify eight of them as Gaia-Enceladus/Sausage stars, which is one of the most massive mergers in our Galaxy's history. We determine their mean age to be \finalage consistent with a merger time of 8-10 Gyr ago. The rest of the stars are possibly associated with Kraken, Thamnos, Sequoia, or another extragalactic progenitor. 
The age determination of ex-situ stars paves the way to more accurately pinning down when the merger events occurred and hence provide tight constraints useful for simulating how these events unfolded.

\end{abstract}




\section{Introduction}
\label{sec:intro}

In the age of large stellar surveys such as \Gaia \citep{GaiaCollaboration2016,GaiaCollaboration2020} and \APOGEE \citep{Majewski2017} it becomes increasingly possible to investigate our Galaxy, the Milky Way, in great detail. Knowledge of the stars' motion on the sky and their chemical compositions provides us with tools to examine where the stars originated.  With precise age determination of single stars, we can contribute to the mapping of the evolution and history of the Galaxy. Using the properties of stars to infer properties of the Galaxy is known as  Galactic archaeology \citep{freeman2002}.

One of the goals of Galactic archaeology is to study the merger history of our Galaxy. Large galaxies like and including the Milky Way are expected to have merged with several dwarf galaxies throughout their lifetimes \citep[see e.g.][]{Helmi1999,Bell2008,Koppelman2019,Kruijssen2020,Elias2020,Naidu2021}. 
Remnants of such a merger in the Milky Way was found by \citet{Helmi2018} using the \Gaia data release 2 \citep[DR2; ][]{GaiaCollaboration2016,GaiaCollaboration2018} and chemical information from \APOGEE \citep[][]{Majewski2017}. The merger was also proposed by  \citet{Belokurov2018} who used \Gaia DR1 dynamical information. Both groups found stars that are kinematically different to the majority of the Milky Way stars and \citet[][]{Helmi2018} furthermore demonstrated that they are also chemically different. The differences in kinematics and chemistry were attributed to them not being born in-situ with the rest of the Milky Way stars but rather ex-situ in a separate dwarf galaxy that had merged with the Milky Way. The dwarf galaxy was named Gaia-Enceladus by \citet{Helmi2018} and Gaia-Sausage by \citet{Belokurov2018} and we will refer to it as \Enc throughout this work.

Several other merger galaxies have been discovered since \Enc, although there are still debate as to which galaxies are unique galaxies and which are part of other already known galaxies  \citep[for a discussion on this see e.g.][]{Helmi2020}. Other such merged galaxies are Sequoia \citep[][]{Myeong2019}, Thamnos \citep[][]{Koppelman2019}, and Kraken \citep[][]{Kruijssen2020} to name a few. Because \Enc is one of the most massive and mainly where the stars in this work appear to originate from (see \autoref{sec:enc}), we will primarily focus on \Enc in this work.

From kinematics the \Enc stars (and many ex-situ stars in general) distinguish themselves from the Milky Way stars by being mainly found in the halo meaning they are not very tightly bound in the Galactic potential and some are even on slightly retrograde orbits. \citet{Koppelman2020} used simulations by \citet[][]{Villalobos2008} of collisions between a Milky Way-like galaxy and a \Enc-like dwarf galaxy with (i) different orbital inclinations and prograde/retrograde configurations and
(ii) different types of progenitors (disky and spherical) to demonstrate that a counter-rotating dwarf spiral galaxy with an in-fall angle of 30$^{\circ}$, relative to the Milky Way disk, would produce a final product with stellar dynamics very close to what we observe for \Enc today. During the collision, the dynamics of the stars in the Milky Way and \Enc were both perturbed but the original counter-rotating signature of \Enc can still be seen in some present day halo stars \citep{Helmi2018}. 
In general, most ex-situ stars will have a different dynamical signature than in-situ stars but because several of the in-situ stars were also perturbed during the merger, the dynamics alone is not enough to distinguish in-situ from ex-situ stars; but chemical compositions can provide us with another diagnostic tool \citep[][]{Baptiste2017}.

\citet[][]{Nissen2010} were one of the first to show that there are two chemically distinct populations in the Milky Way halo, which was one of the first indications of a different galaxy being embedded in our Milky Way. 
The chemical evolution of the interstellar medium -- and thereby the surface abundance of newborn stars -- is predominantly governed by the rate and types of supernovae explosions, which in turn is controlled by the star formation rate (SFR) of the host galaxy. 
Simply put, there are two main types of supernovas produced by either massive stars ($\gtrsim$ 8 M$_\odot$) or low-mass stars. 
Massive stars have shorter life times and explode as core-collapse supernovas or \emph{type II supernovae} (SNII).
This kind of SNII produces large amounts of $\alpha$-elements such as O, Mg, Si, S, Ca and Ti as well as other elements such as Na and Al. Once the longer lived lower mass stars have had time to evolve to white dwarfs they can -- if they are in a binary system --  collide with another white dwarf companion to produce a \emph{type supernova Ia} \citep[SNIa;][]{whelan1973,iben1991,kromer2015}. The SNIa releases very small amounts of $\alpha$-elements but large amounts of for example iron. When the occurrence rate of SNIa increases the $\alpha$-element content of the interstellar medium stays mostly constant while the [Fe/H] content increases. As \alphafe depends on the amount of iron released in the interstellar medium, \alphafe will decrease when the SNIa sets in. This produces a bend in the \alphafe vs [Fe/H] relation with \alphafe being constant early in the galaxy's lifetime and later decreasing as [Fe/H] increases. Where this bend or "knee" occurs depends on the mass and SFR in the galaxy \citep[see e.g.][]{Howell2014,Helmi2018}. 
Massive galaxies have more gas than less massive galaxies and can form several generations of stars throughout its life. \citet{dave2008} show a tight relation between galaxy mass and SFR up til z${\sim}2$. Smaller dwarf galaxies that have less gas have a smaller SFR and the onset of Supernova Type Ia explosions happens at a lower [Fe/H] abundance. The stars from smaller galaxies, therefore, have a lower [Fe/H] abundance at the same \alphafe abundance compared to stars in a more massive galaxy. In the literature, this interpretation of the behaviour of observed abundance ratios such as \alphafe vs [Fe/H] in different galactic systems is known as the time-delay model \citep{tinsley1979,matteucci1986,matteucci2012}. 
Other elements such as Al are also released in SNII explosions. The production of Al depends on C and N and increases with metallicity until SNIa sets in and the [Al/Fe] decreases. This makes [Al/Fe] another good tracer for what kind of galaxy a star was formed in \citep{Hawkins2015,Das2020,Buder2021}.  Apart from iron, Mn is also mainly produced in SNIa and the relation between [Mg/Mn] and [Al/Fe] provides a further diagnostic to distinguish between stars born in galaxies of different mass (for a more detailed description of this see for example \citet[][]{Das2020} and \citet[][]{Buder2021} and references therein).  
Because the dwarf galaxies and the Milky Way are very different in mass \citep[e.g. ratio ${\sim}1$:$4$ or M$_{\mathrm{\Enc}}{\sim}10^{10}\mathrm{M}_\odot$ for \Enc;][]{Belokurov2018,Helmi2018,Feuillet2020,Naidu2021} the difference in the chemical abundance of the stars can be used to identify the stars originating from another galaxy than the Milky Way. 

When the ex-situ stars have been identified, the next step is to determine their ages. Low-to-intermediate-mass evolved \RG (RG) stars are particularly suited for this purpose because they are bright and thus can be seen at large distances and most importantly they exhibit solar-like oscillations and hence can be studied using asteroseismology \citep[][and references therein]{Aerts2010book}. These oscillation patterns change depending on the size and density of the stars, which makes it possible to pin down the stellar properties to an extraordinary precision (more on this in \autoref{sec:agedetermination}). 
With an asteroseismic analysis it is possible to determine the stellar age to better than 25\%  \citep{Casagrande2016,Aguirre2018}. 
The ages of the stars can be used to estimate when the merger happened because we may assume that all star formation took place in the galaxy before it was fully disrupted, and also because during mergers gas is stripped off, removing the galaxy from the fuel to produce stars.

As we mainly focus on \Enc in this work, we briefly introduce some studies on this galaxy.
The time of the \Enc merger has been debated since its discovery but studies using isochrone fitting or galaxy modeling have shown that the merger ended some 8-10 Gyr ago \citep[e.g.][]{Belokurov2018,Helmi2018,Gallart2019,Grunblatt2021}. The merger time can be estimated from isochrone fitting based on the assumption that the youngest stars are formed shortly before full disruption of their parent galaxy.
There are several additional interesting works worth mentioning in the context of the Galactic archaeology of the \Enc merger. One is the study of \nuindi by \citet{Chaplin2020}. \nuindi is a metal-poor subgiant star, which they show was born in-situ but was kinematically heated by the merger of \Enc. Its age has been determined to be  11.0 $\pm$ 0.7 (stat) $\pm$ 0.8 (sys) Gyr indicating that the merger must have happened $11.6$ Gyr ago at the earliest. Although this is one of the best age determinations of a metal-poor star that we have, the fact that it is a single in-situ star limits the information it can provide for the merger time. 
\citet{Kruijssen2020} found globular clusters expected to belong to \Enc and used a Neural Network trained on cosmological simulations to estimate an accretion time of 9.1$\pm$0.7 Gyr ago. Although globular clusters are very useful for age determination, asteroseismology provides us with additional information that can yield much more precise ages and compared to the Neural Network in for example \citet[][]{Kruijssen2020}, asteroseismology also provides a direct estimate. 
Such an asteroseismic analysis was made by \citet[][hereafter \Montalban]{Montalban2021} who found seven red giant branch stars, which they classified as belonging to \Enc. A comparison of stellar ages and selection criteria between our work and theirs is presented in \autoref{sec:enc}.

In this paper, we explore the dynamics and chemical properties of \RG stars in order to identify which of them originates from other galaxies than the Milky Way. In \autoref{sec:data} we provide an overview of our sample. In \autoref{sec:selection} we demonstrate how we select the sample of ex-situ stars. For these stars we use the Bayesian statistics software \BASTA to determine their age given asteroseismic parameters, all of which is further described in \autoref{sec:agedetermination}. The results and a discussion on the ages and the origin of the ex-situ stars is presented in \autoref{sec:discussion} and a conclusion is given in \autoref{sec:conclusion}.


\section{Data sample}
\label{sec:data}
\begin{figure}
	\centering
	\includegraphics[width=8.8cm]{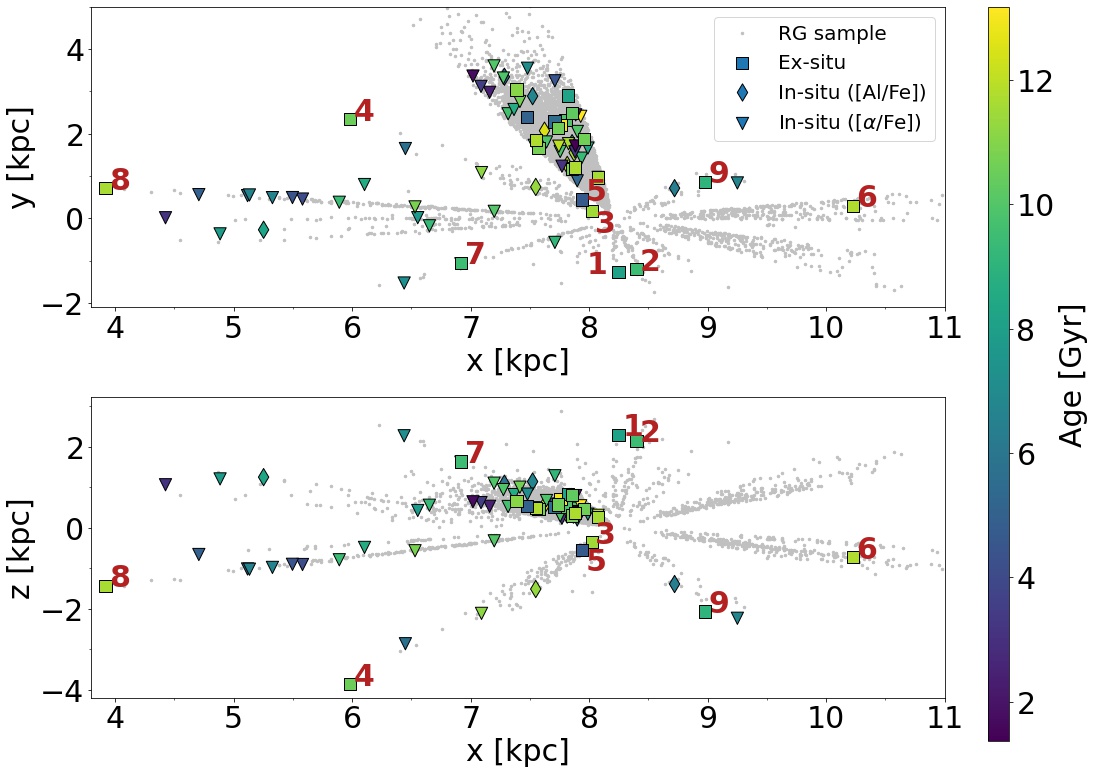}
	\caption{Position of the stellar sample in Galactocentric cartesian coordinates. Grey dots are the full \RG star sample while the coloured points mark the stars of interest in this work, coloured by their stellar age. The different symbols denote different selection criteria (for more information see \autoref{sec:selection}) with the squares being the ex-situ stars. Indices are only added to the ex-situ from the \kt mission in the interest of clarity.}
	\label{fig:position}
\end{figure}

Our full sample of stars consists of nearly 12,000 \RG stars observed by the \emph{Kepler Space Telescope} \citep[the original \kepler mission and/or the succeeding \kt mission; ][]{Borucki2010,Koch2010,Howell2014,Stello2017} and is compiled from the \RG stars from \citet{Yu2018} (\kepler) and the stars from \citet[][]{Stello2017} (K2 campaign 1) and \citet[][]{Zinn2021} (K2 campaigns other than 1). The galactic positions of the stars in our sample can be seen in \autoref{fig:position}. 
This catalogue of \RG stars is cross-matched with data from \Gaia early Data Release 3 \citep[eDR3; ][]{GaiaCollaboration2016,GaiaCollaboration2020}, Apache Point Observatory Galactic Evolution Experiment's DR16 \citep[\APOGEE; ][]{Majewski2017,Ahumada2020} and the Two Micron All Sky Survey \citep[2MASS; ][]{Skrutskie2006}. 

Our sample was pruned by selecting stars for which we have asteroseismic parameters (\numax and \dnu, see  \autoref{sec:asteroseismology}), 6D phase space parameters from \Gaia eDR3, metallicity and temperature from \APOGEE, and photometry from three filters ($J$,$H$, and $K_s$) from \TMASS. Furthermore, we only kept stars with \texttt{phot-bp-rb-excess-factor}$\leq1.27$ and a \texttt{RUWE}$<1.4$ to limit the stars with poor photometric and astrometric data. The \TMASS quality must be \texttt{AAA} and for APOGEE data we use stars without bad flags in \texttt{MG$\_$FE$\_$FLAG, SI$\_$FE$\_$FLAG, FE$\_$H$\_$FLAG, BAD$\_$PIXELS, VERY$\_$BRIGHT$\_$NEIGHBOR, LOW$\_$SNR} and \texttt{SUSPECT$\_$RV$\_$COMBINATION} as well as \texttt{TEFF$\_$BAD, METALS$\_$BAD, STAR$\_$WARN, VSINI$\_$BAD} and \texttt{STAR$\_$BAD}.

It has been shown that the \Gaia parallaxes have a systematic off-set of 17 $\mu$as \citep{Lindegren2020}, which we correct for. 
To account for potential underestimated uncertainties of \APOGEE metallicities, we use a lower limit of $0.1$ dex in [Fe/H] and [$\alpha$/Fe] uncertainty when fitting the stars but not when plotting them. The orbital kinematics of the stars are calculated using  \textsc{Galpy}\footnote{\url{http://github.com/jobovy/galpy}} fast orbit estimation algorithm \citep{Bovy2015,Mackereth2018} and the \texttt{McMillan2017} potential \citep{McMillan2017} assuming (U, V, W) = ($11.1, 12.24, 7.25$) km~s$^{-1}$  \citep{Schonrich2010}, v$_\mathrm{LSR}$ = $221$ km~s$^{-1}$  for the local standard of rest and the Sun's distance to the galactic centre of $8.2$ kpc \citep{McMillan2017}. 
The uncertainties on the dynamical quantities were calculated using a bootstrap method by randomly drawing a sample of phase-space quantities based on the uncertainties and covariance matrix provided for the \Gaia parameters. Each quantity is drawn 10.000 times and from these, the median and 16th and 84th quantiles were calculated and used as the value and corresponding uncertainty.

\section{Selecting Ex-situ stars}
\label{sec:selection}
Distinguishing ex-situ stars from in-situ stars is not a trivial process. The sample might be contaminated by in-situ stars that were for example heated by mergers and some ex-situ stars might have orbits that are not distinguishable from the in-situ stars. There is also a possibility of contamination of in-situ stars due to for example, ill-determined observed properties, or underestimated uncertainties. To combat this and get as pure a sample of ex-situ stars as possible, we make a stringent selection. This robust selection is based on a cut in dynamical space and two cuts in different chemical spaces (see \autoref{tab:selection}). As mentioned in \autoref{sec:intro}, the Milky Way has merged with several galaxies in the past, meaning it is possible that the ex-situ stars we find are not all from the same merger remnant (see \autoref{sec:exsitu}). We therefore distinguish between ex-situ and \Enc stars throughout this work.

\begin{figure}
	\centering
	\includegraphics[width=8.8cm]{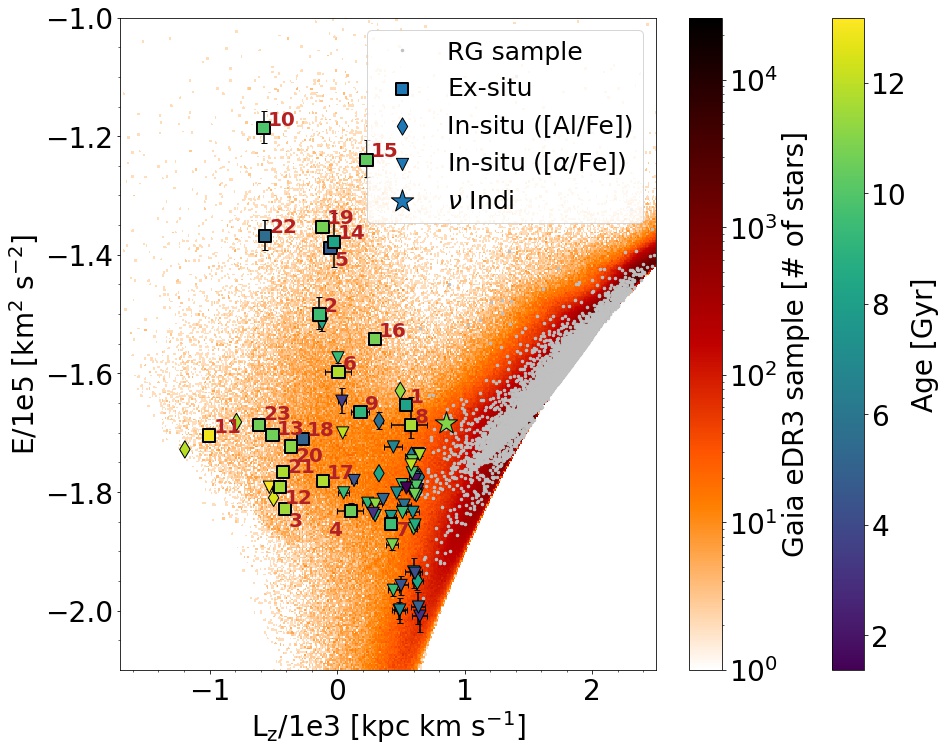}
	\caption{Distribution of stars in \Lz-\En space.  The gray points denote the full sample of RG stars, the different symbols denote the classification of our selected stars as described in \autoref{sec:selection}, and the orange background stars are the full \Gaia eDR3 sample. The star symbol is \nuindi with age from \citet{Chaplin2020}. All ex-situ stars are marked with the indices corresponding to those in \autoref{tab:All stars}. }
	\label{fig:dynamics}
\end{figure}

The first selection criterion is in dynamic space meaning in energy (\En) and angular momentum (\Lz). This has become a common method of identifying ex-situ and \Enc stars and was partly also how the \Enc stars were discovered in \citet{Helmi2018}. We select stars with small and negative angular momentum (\Lz$<0.65$e$3$ kpc km s$^{-1}$). Negative angular momentum means that the stars are counter-rotating compared to the rest of the Milky Way stars, that have positive angular momenta and thus it is an indication that these stars are not formed in-situ. In \autoref{fig:dynamics}, the full \RG star sample is in gray and the stars from this dynamic selection are marked by coloured symbols. The full sample of \Gaia eDR 3 stars (with \texttt{RUWE}$<1.4$) are shown in orange shade for easier comparison to other works. The selection leaves us with $78$ stars and we determine the ages for all of them.

\begin{figure}
	\centering
	\includegraphics[width=8.8cm]{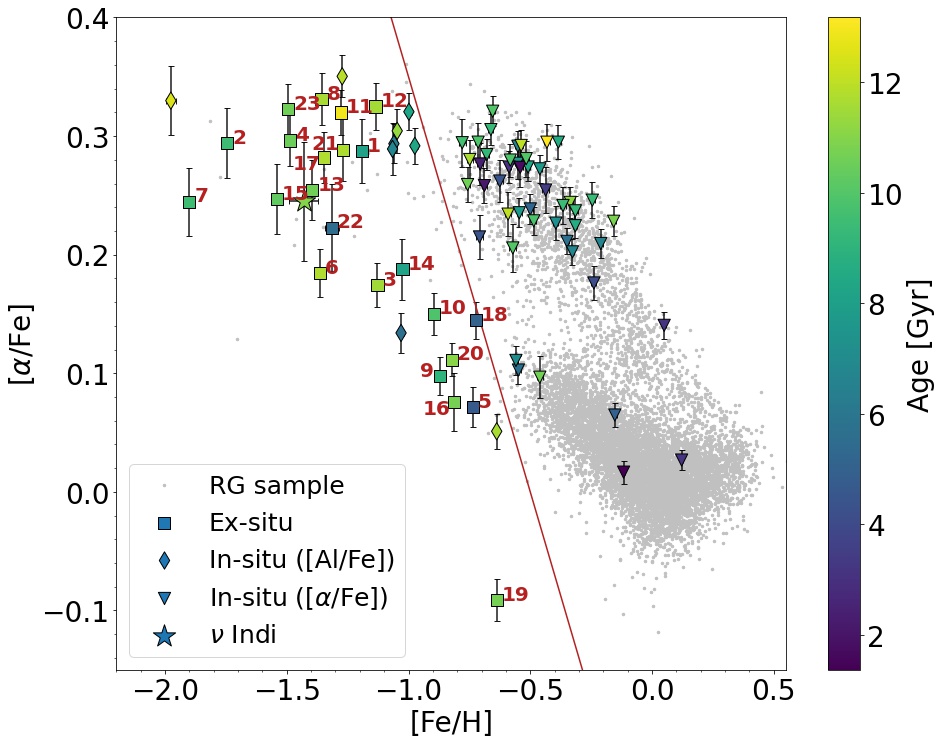}
	\caption{[$\alpha$/Fe] vs metallicity abundance. Markers, colours and indices are the same as in \autoref{fig:dynamics}. The red line marks our selection criterion (see \autoref{tab:selection}). The metallicity for \nuindi is from \citet[][]{Chaplin2020}.}
	\label{fig:alpha}
\end{figure}

The dynamics cut is not sufficient to fully exclude the in-situ stars. Therefore, we make two additional cuts in chemical space (see \autoref{tab:selection}). The first is in [$\alpha$/Fe] vs [Fe/H] space as shown in \autoref{fig:alpha}. Here, [$\alpha$/Fe] is defined as $\frac{1}{2}$([Mg/Fe]+[Si/Fe]) following \citet{Salaris2018}.
In this plot, there are two distinct populations, one that correlates with the in-situ stars at high [Fe/H] abundance and one at lower [Fe/H], which matches with our expectations of them being ex-situ stars. We make a division between these two populations with line at [$\alpha$/Fe]$>-0.55$[Fe/H]$-0.25$ (red line in \autoref{fig:alpha}). 
All stars from the dynamical selection that fall above this criteria are marked with triangles and are denoted as in-situ stars.  With this cut we sort out $48$ of the $78$ stars and classify them as in-situ stars. 

\begin{figure}
	\centering
	\includegraphics[width=8.8cm]{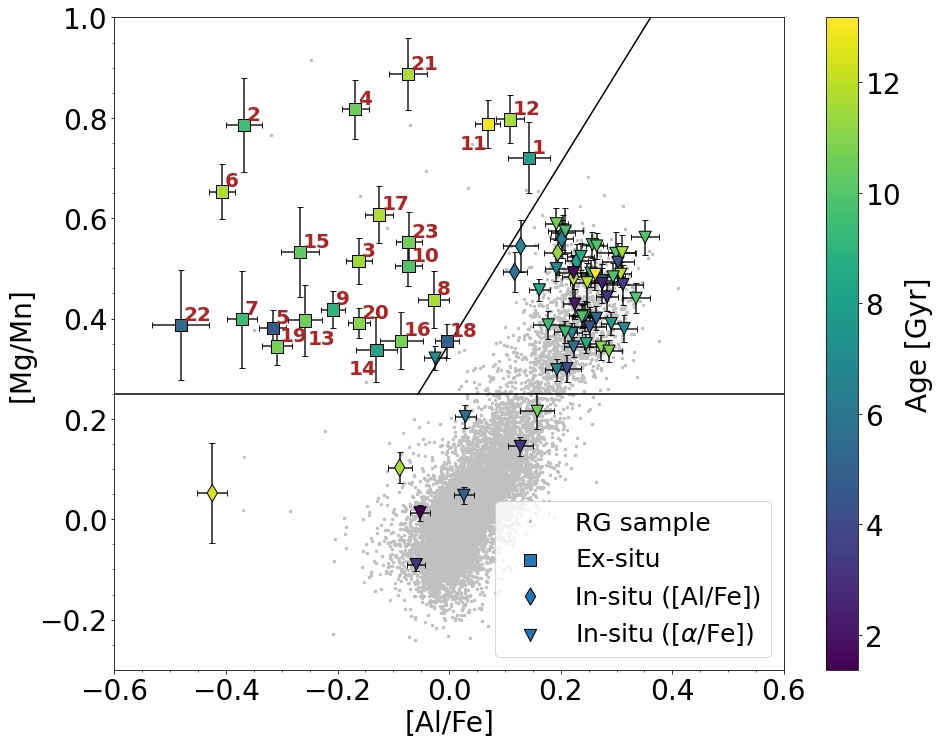}
	\caption{[Mg/Mn] vs [Al/Fe] abundance with lines from \citet{Horta2021} (see also \autoref{tab:selection}). Markers, colours and indices are the same as in \autoref{fig:dynamics}.}
	\label{fig:mgmn}
\end{figure}

Lastly a cut is made in [Mg/Mn] vs [Al/Fe] space based on the chemical evolution of galaxies described in \autoref{sec:intro}. The lines dividing the populations are based on \citet{Horta2021} (see \autoref{fig:mgmn} and \autoref{tab:selection}). 
Here, stars in the upper left corner are regarded as ex-situ stars. Stars that are removed from the ex-situ sample due to this cut are marked with diamonds unless they were already removed due to the [$\alpha$/Fe] selection, in which case they remain denoted with triangles. This last cut removes $7$ additional stars and deem them in-situ stars. 

\begin{table}
\caption{Selection criteria for ex-situ star classification}
\label{tab:selection}
\begin{center}
\begin{tabular}{ r l l }
& This work & \Montalban\\
\hline
 1. & \Lz$<0.65\times10^3$ kpc km s$^{-1}$ & e$\,>0.7$\\ 
 2. & [$\alpha$/Fe]$\,<-\,0.55\,$[Fe/H]$\,-\,0.25$ &  [Mg/Fe]$\,<-0.2\,$[Fe/H]$\,+\,0.05$\\  
 3. & [Mg/Mn]$\,>1.8\,$[Al/Fe]$\,+\,0.35$ & \\
   & [Mg/Mn]$\,>0.25$ & 
\end{tabular}
\end{center}
\end{table}

In the end, only stars that are in the ex-situ sample in all three parameter spaces are classified as ex-situ stars and denoted with squares in all figures. We find a total of $23$ likely ex-situ stars in our sample. These stars are marked with indices in all figures for easier identification. The relation between indices and KIC and EPIC ID's can be seen in \autoref{tab:All stars}. We note that of the $78$ stars purely selected in dynamical space only ${\sim}1/3$ are classified as ex-situ stars. This shows that a simple selection in \Lz can result in significant amounts of contamination. 

\begin{figure*}
	\centering
	\includegraphics[width=17.6cm]{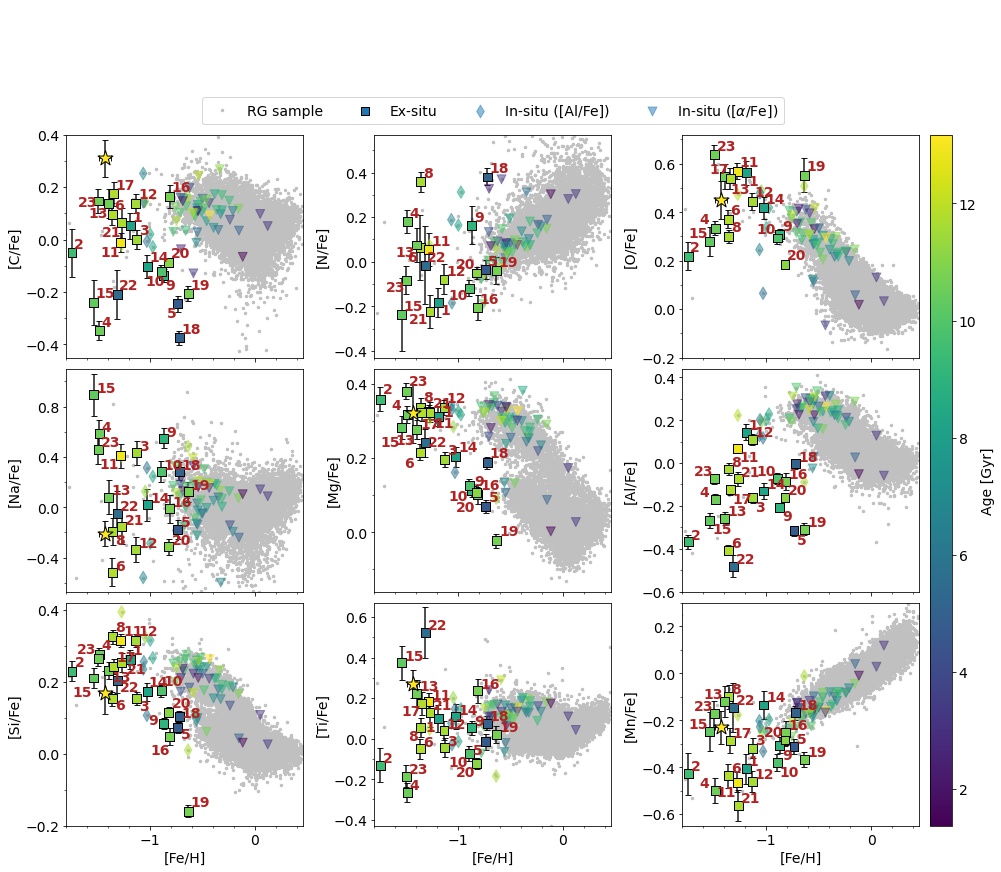}
	\caption{Different metallicities for our sample of \RG stars. Not all abundances were available for all stars. The colour and indices are the same as previous figures. All figures has [Fe/H] in the x-axis with the same scale.}
	\label{fig:multi_metallicity}
\end{figure*}

In \autoref{fig:multi_metallicity}, we show the stars in a number of different chemical spaces. Common for all is that the stars we select as ex-situ appear to be a different population than the gray points we expect to be in-situ stars. This strengthens our assumption that these stars indeed are formed in different galaxies. 

Five of the classified in-situ stars are on retrograde orbits \linebreak (\Lz$<0$ kpc km s$^{-1}$ in \autoref{fig:dynamics}) which is peculiar for in-situ stars. They are removed from the ex-situ sample based on chemistry but a follow-up spectroscopic survey of these stars could clarify their classifications. In this work, we keep them as in-situ and note it is stars \#51, 53, 64, 67, and 77 in \autoref{fig:Agesx3} and \autoref{tab:All stars}.

\section{Age Determination}
\label{sec:agedetermination}
In this section we describe the theory behind asteroseismology (\autoref{sec:asteroseismology}) and how it is used for age determination. In \autoref{sec:basta} and \autoref{sec:grid} we present the code and grid used for the fitting and in \autoref{sec:fit} we describe the details of the stars that are fitted. 
\subsection{Asteroseismology}
\label{sec:asteroseismology}
As mentioned in \autoref{sec:intro}, we use constraints from asteroseismology to determine the stellar ages of our sample.
Asteroseismology is the study of how the stars oscillate and vibrate.
The great advantage of asteroseismology is that the pattern in the power spectrum of the photometric time series is structurally identical for all solar-like stars. The power spectrum pattern consists of frequencies ($\nu$) that have a very regular comb structure (at least for slowly rotating stars) and follow a Gaussian-like shape in power as a function of frequency \citep[see e.g.][]{Aerts2010book,Basu2016}. As solar-like oscillators shows this general structure, we can decompose this pattern into two global asteroseismic parameters: the large frequency separation (\dnu) and the frequency at maximum power (\numax). As the name suggests \numax is the frequency where the Gaussian-like distribution peaks and the large frequency separation (\dnu) is the difference in frequency between oscillations of the same angular degrees ($l$) but of consecutive radial order ($n$) hence
\begin{align}
   \Delta \nu_l(n)=\nu_{n,l}-\nu_{n-1,l}.
\end{align} 

\citet{Ulrich1986} and \citet{JCD1988} have analytically shown that \dnu scales with radius ($R$) and mass ($M$) of the star in the following manner:
\begin{align}
\label{eq:dnu}
    \frac{\Delta \nu}{\Delta \nu_\odot} \simeq \left( \frac{M}{\mathrm{M}_\odot}\right)^{1/2} \left( \frac{R}{\mathrm{R}_\odot}\right)^{-3/2}.
\end{align}

\noindent Furthermore, \citet{Brown1991}, \citet{Kjeldsen1995} and \citet{Bedding2003} found a semi-empirical scaling relation based on the frequency of maximum power, which scale with mass, radius, and effective surface temperature (\teff) of the star
\begin{align}
\label{eq:numax}
    \frac{\nu_\mathrm{max}}{\nu_{\mathrm{max},\odot}} \simeq \left( \frac{M}{\mathrm{M}_\odot}\right) \left( \frac{R}{\mathrm{R}_\odot}\right)^{-2} \left( \frac{{T}_{\mathrm{eff}}}{\mathrm{T}_{\mathrm{eff},\odot}}\right)^{-1/2}.
\end{align}

\noindent Combining \autoref{eq:dnu} and \autoref{eq:numax}, we can estimate the mass and radius of the star only based on the asteroseismic parameters and the surface temperature.
The mass and radius are determined as
\begin{align}
    \left( \frac{M}{\mathrm{M}_\odot}\right) \simeq \left(\frac{\nu_\mathrm{max}}{\nu_{\mathrm{max},\odot}}\right)^3  \left( \frac{\Delta \nu}{\Delta \nu_\odot} \right)^{-4} \left( \frac{T_{\mathrm{eff}}}{\mathrm{T}_{\mathrm{eff},\odot}}\right)^{3/2}
\end{align}

\noindent and

\begin{align}
\label{eq:radius}
    \left( \frac{R}{\mathrm{R}_\odot}\right) \simeq \left(\frac{\nu_\mathrm{max}}{\nu_{\mathrm{max},\odot}}\right)  \left( \frac{\Delta \nu}{\Delta \nu_\odot} \right)^{-2} \left( \frac{T_{\mathrm{eff}}}{\mathrm{T}_{\mathrm{eff},\odot}}\right)^{1/2}.
\end{align}

This way the asteroseismic parameters are directly related to the physical properties of the star. The scaling relations between the observables \dnu and \numax (along with \teff) and the stellar properties like the mass and radius of the star are one of the great successes of asteroseismology. 
Asteroseismology narrows down the likely parameter space and provides tight constraints valuable for age determination.

The resolution of the power spectra depends on for example the length of time series data, and if the resolution is good enough individual mode frequencies can be extracted. This is especially the case for many \kepler stars. Individual frequencies can greatly improve the precision of the fits because they can constrain the stellar properties even further than \dnu and \numax. 

For extraction and identification of the frequencies we use the python package \PBJam \citep{PBJam}. The frequencies have been corrected for the Doppler shift caused by the stellar line-of-sight movement in accordance with \citet[][]{Davies2014}.
If individual modes are not available we use the \numax and \dnu values from the \SYD pipeline \citep{Huber2009} and an AI-detector for the classification of the evolutionary state \citep{Yu2018,Hon2018,Zinn2020}. 
For many of the \kt stars an AI-vetter was used to determine the quality of the \dnu measurements (Reyes et al. 2021, submitted). This algorithm provides a metric for the relation of the purity and completeness and we only accept stars with this number larger than 0.6,  which corresponds to a purity of ${\sim}97$\% and a completeness of ${\sim}93$\%. 
To account for systematic uncertainties of the \kt stars, we added a percentage of \dnu and \numax to the uncertainties based on the standard deviation of the correction factors of different pipelines in \citet[][their Table 3]{Zinn2021}. We estimated the systematic \numax uncertainty to be $0.3\%$ for stars that are known to be in the red giant branch (RGB) phase and $0.7\%$ for those in the red clump (RC) (or unknown) phase. The systematic \dnu uncertainty is estimated to be $0.3\%$ for RGB and $0.4\%$ for RC (or unknown) phase. 

For the stars that only had \dnu and \numax we made a consistency check with the APOKASC sample \citep{Pinsonneault2018} and the K2 GAP DR3 sample \citep{Zinn2021} for the \kepler and \kt stars, respectively. All stars in this work that were available in one of the two studies had consistent \dnu and \numax values within $\pm1\sigma$ of the respective studies, except star \#6. It appears the \SYD pipeline overestimates the \numax of this star with ${\sim}2\mu$Hz compared to other pipelines. Computing the stellar properties for star \#6 with the K2 GAP DR3 \numax also resulted in a more consistent fit with respect to the other input parameters and we chose to use this value of \numax for this star. For all other stars we keep the values from the \SYD pipeline.

\subsection{\BASTA}
\label{sec:basta}
To determine the stellar parameters we use the BAyesian STellar Algorithm \citep[\BASTA; ][]{Aguirre2015,Aguirre2017,Aguirre2021}. The algorithm uses a pre-calculated grid of stellar tracks and Bayesian statistics to find the best fitting stellar parameters for each star. \BASTA allows for prior probability distributions to be taking into account when calculating the fits. We use the Salpeter initial mass function \citep{Salpeter1955} to account for the expected mass distribution of stars, favouring low-mass stars as the most abundant. We use the two-term surface correction described in \citet{Ball2014}. Additionally, we include an upper limit on the stellar ages of 15 Gyr. This is done to avoid nonphysical solutions for stars older than the age of the universe. Despite the solutions not being physical at above the age of the universe (13.7 Gyr), they can still hold statistical significance and we do therefore not truncate the solutions at 13.7 Gyr but allow them to stretch to 15 Gyr. For the remaining parameters we use uniform priors.

\subsection{The grid}
\label{sec:grid}
As mentioned above, \BASTA uses a grid of stellar models to fit the stars. 
We build a quasi-random sampled \citep[sobol;][]{sobo67} grid with ${\sim}8000$ evolutionary tracks of stellar models using the Garching Stellar Evolution Code \citep[\Garstec]{weiss2008}. The mass range of the grid is between $0.7$ and $2.0$ M$_\odot$, the initial metallicity [Fe/H]$_{\mathrm{ini}}$  varies between $-2.4$ and $0.1$ dex, the $\alpha$-enhancement ranged from $-0.2$ to $0.6$ dex in steps of $0.1$ dex. The mass loss ($\eta$) ranges from $0.0$ to $0.3$ following the \citet{Reimers1977} formalism. Convection in the models are parameterised using mixing-length theory \citep{Bohm1958} and the mixing-length parameter is kept constant at the solar-calibrated value of $1.789$ as determined from a standard solar model calibration to the  \citet[][]{Asplund2009} abundances. The initial helium is assumed to be 0.248 \citep{Fields2020} and the helium-to-metal ratio is $\Delta Y/\Delta Z=1.4$. 
The stellar models are evolved from pre-main sequence to the beginning of the RGB whereafter the models are saved and frequencies computed along the RGB, through the RC and all the way to the asymptotic giant branch (AGB). This provides us with a very fine grid of stellar models in the RGB and RC phase. For all models the radial oscillation modes ($l=0$) are computed using the Aarhus adiabatic oscillation package \citep[\textsc{adipls};][]{JCD2008}.

\subsection{Fits}
\label{sec:fit}
To determine the ages of the stars we use the observed parameters \alphafe, [Fe/H], \teff, asteroseismic values (individual frequencies or \dnu and \numax, as well as evolutionary phase when available) and distance.
We use the parallax and three photometric filters when fitting the distances to the stars (\TMASS filters $J$, $H$, and $K_s$) as well as the dust map from \citet{Green2019} for computing the extinction.
These parameters are mapped to our grid of stellar models to find the best match. 

When we fit the stars, different constraints can point towards slightly different solutions. Especially the constraint of the parallax and the photometric colours can be in tension with the asteroseismic solutions. This is typically due to inaccurate or incorrect determination of one or more of the parameters, underestimates of the uncertainties, or models inaccuracies. As a test of robustness, we fitted the stars in three different ways: (i) by fitting all the parameters mentioned above, (ii) by fitting everything excluding the distance (parallax and photometric filters), and (iii) fitting everything excluding the asteroseismic values. Since the asteroseismic values and distance both give a direct measure of the radius of the stars (see \autoref{eq:radius}), we compare the solution for the radius of the three fitting methods. In the cases where the calculated median radius of the fit with all parameters (i) agrees within $\pm1\sigma$ with the solution of both other fits (ii and iii), we accept the solution (i) as robust and it is chosen because adding additional information improves the uncertainties of the solution. This check is in addition to a manual inspection of the probability distributions, to insure the solution for example does not interfere with the edge of the grid. 

For some stars, the three fits did not match up and gave two or even three very different solutions. The fit with all the parameters (i) would then artificially be located between the two solutions of (ii) and (iii) or pushed to the edge of the two. In these cases, we will chose to trust  case (ii), that is the fits including asteroseismic constraints  but excluding the constraints from photometry and astrometry. We choose the asteroseismic solution as many aspects can influence the distance fit such as poorly determined parallax, colours or dust map to name some. Choosing the asteroseismic solution over the full fit mainly affects the uncertainties of the solution. Because our selection of the stars is based on \En and \Lz (which is calculated with the parallax) we check that the selection is not affected if we now adopted the distance output form \BASTA in the selection (by recalculating E and Lz).  While some of the stars did move slightly in \En and \Lz space, they where all still part of our selection. 
In \autoref{tab:All stars}, we note which stars have been fitted with what method and whether we used individual frequencies or \dnu and \numax. Ten of the $23$ ex-situ stars and $19$ of the $55$ in-situ  stars have been fitted without the distance.

\begin{figure}
	\centering
	\includegraphics[width=8.8cm]{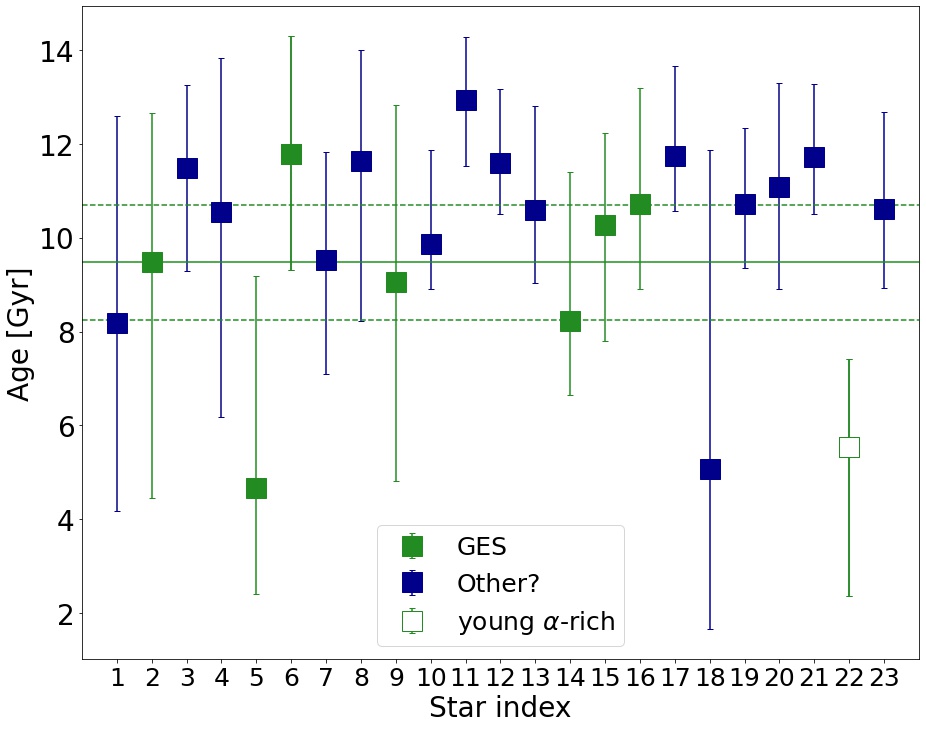}
	\caption{Ages of the ex-situ stars with uncertainties. Indices match the ones used in other figures. Green symbols are the ones we classify as \Enc stars and blue stars are the ones that might originate from other mergers. The hollow symbol for star \#22 marks that this might be a young $\alpha$-rich star. The full horizontal line indicates the mean age of the green points (the \Enc stars) and the dashed lines indicate the standard deviation.} 
	\label{fig:Ages}
\end{figure}

The determined ages of the stars are shown as colour coding in \autoref{fig:position}, \ref{fig:dynamics}, \ref{fig:alpha}, \ref{fig:mgmn}, and \ref{fig:multi_metallicity}, the ages of the ex-situ stars with uncertainties are shown in \autoref{fig:Ages} and for all $78$ stars in \autoref{fig:Agesx3} and \autoref{tab:All stars}.

\section{Results and Discussion}
\label{sec:discussion}	
In this section we present the results of the age determinations. In \autoref{sec:peculiar}, we discuss some stars with peculiar solutions and how we treat them. In \autoref{sec:exsitu} we discuss the origin of the ex-situ stars and how the ages can give us an estimate of the upper limits of the merger times.

\subsection{Peculiar stars}
\label{sec:peculiar}
A peculiar case among the ex-situ stars is star \#22, which is very young compared to the rest of the ex-situ stars (see \autoref{fig:Ages} or \autoref{tab:All stars}). Although it is likely a \Enc star (see \autoref{sec:enc}) it appears too young because \Enc does not have any recent star formation. This is the \kepler star KIC 8694070, which has individual mode frequencies available. The solutions from the fits with all parameters and those without distance did not agree for this star (see discussion in \autoref{sec:fit}), and we defaulted to using the fit without the distance in this case. The asteroseismic fit is good and we trust the solution, which could mean this is a young $\alpha$-rich star.
As described above, the asteroseismic parameters are used to find a mass and a radius of the stars, which can then be compared to an age with stellar evolutionary tracks to find the best correlation. If the star was part of a multistellar system where mass transfer have occurred, the current mass could exceed the initial mass of the star and make the star appear younger than its actual age. This is normally known as blue stragglers or in the case of old \RG stars as young $\alpha$-rich stars \citep{Martig2015,Zhang2021}. This means that this star is likely not actually this young and we exclude it from any determination of mean ages in later calculations.

Star \#4 has a \dnu$=0.64\pm0.12\mu$Hz and our grid is limited to \linebreak \dnu$>0.6\mu$Hz, which means this star is very close to the edge of the grid. This star is also a \kt stars with a short timeseries and a very small \numax$=4.79\pm0.62\mu$Hz. Asteroseismic values this low can be very hard to measure for stars with short timeseries and we have therefore chosen to not fit \dnu and \numax for this star, as we do not find them reliable. With more data from for example the \TESS satellite \citep[][]{Ricker2015} it is possible that the asteroseismology could be improved and used for fitting in the future. 

\subsection{Ex-situ}
\label{sec:exsitu}
\begin{figure}
	\centering
	\includegraphics[width=8.8cm]{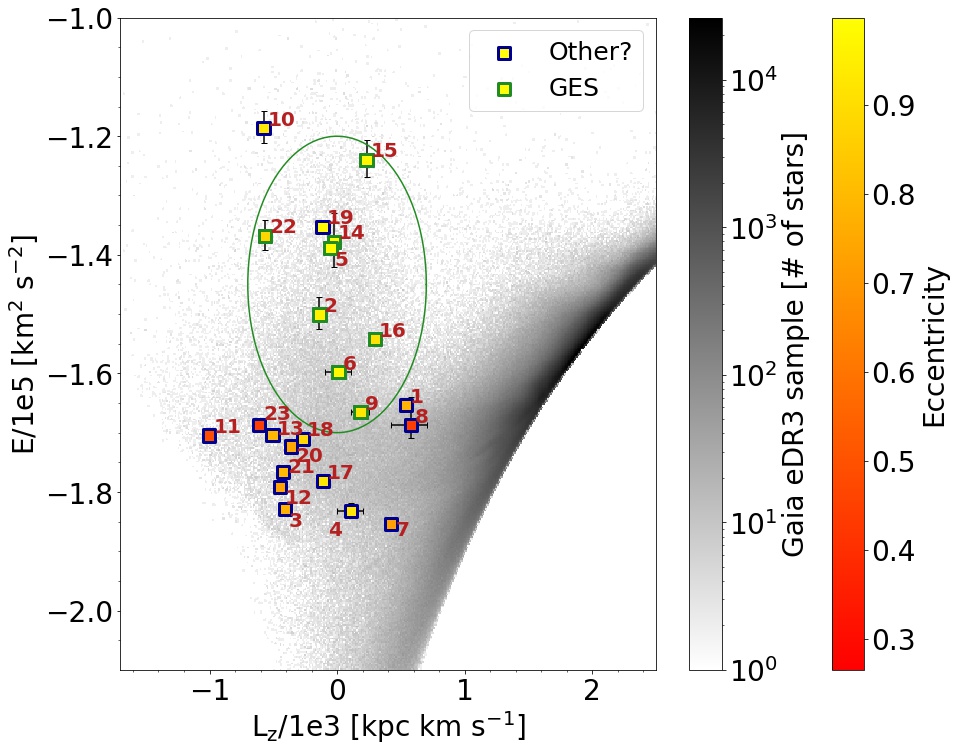}
	\caption{Similar to \autoref{fig:dynamics} but with only ex-situ stars colour coded according to eccentricity of the orbits. In gray is the full sample of \Gaia eDR3 stars. The green outline denotes stars that we classify as \Enc and the blue as possibly originating from other merger galaxies. The green circle shows how we select the stars that are classifies as \Enc.}
	\label{fig:Dynamics_enc}
\end{figure}

\begin{figure}
	\centering
	\includegraphics[width=8.8cm]{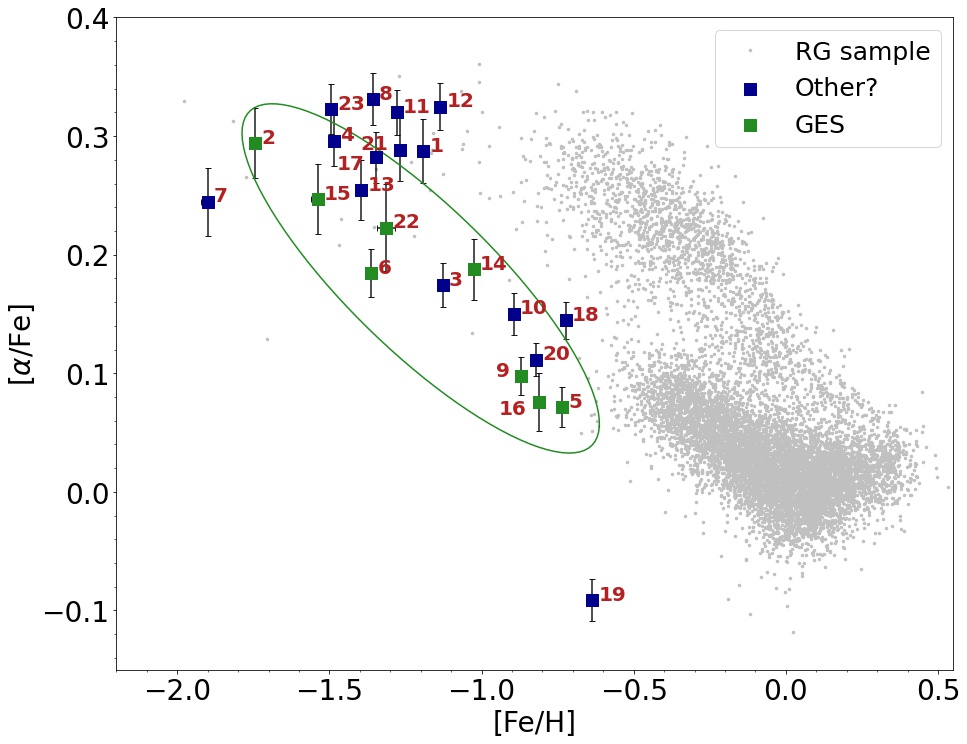}
	\caption{Similar to \autoref{fig:alpha} but only ex-situ stars coloured according to our classification as \Enc (green) or possibly originating from other mergers (blue)}
	\label{fig:alpha_fe_enc}
\end{figure}

Through our selection criteria we have determined which stars are ex-situ stars in our sample. Associating them with a specific merged galaxy is more tricky. There are several studies denoting stars and globular clusters to galaxy remnants such as \citet[][]{Massari2019,Koppelman2019,Kruijssen2020,Naidu2021}. The issue with comparison between different works are for example that they use different potentials to calculate the dynamics, making it hard to do direct comparison. The classification within the same potential is also not always clear and remnant stars from different galaxies overlap in both kinematic and chemical space making it nearly impossible to distinguish. 
In \autoref{fig:Dynamics_enc}, we show the ex-situ stars similar to \autoref{fig:dynamics} but colour coded according to eccentricity ($e$). It is generally accepted that \Enc stars have larger \En than stars with similar \Lz and large $e$ but the exact boundaries are not clear. The difference in chemistry (see \autoref{fig:alpha_fe_enc}) between our ex-situ stars suggests that these are not all of the same origin. It is possible that all these are indeed \Enc stars but due to the large spread in dynamics and chemistry we have chosen to make a very conservative classification of the \Enc stars which is described in \autoref{sec:enc}. In \autoref{sec:kraken} and \autoref{sec:other}, we discuss other possible origins of the remaining stars. 

As mentioned in \autoref{sec:intro}, the accreted galaxies are disrupted during the merger, which quenches star formation. The mean age of the stars originating in a galaxy can, therefore, be interpreted as an upper limit for when the galaxy was disrupted during the merger. We present such mean values below. The mean age is calculated with a bootstrap method with $1000$ iterations.

In general, all the stars are older than $8$ Gyr (except for star \#22), which is in good agreement with expected ages for accreted stars \citep[see e.g.][]{Kruijssen2020}.


\subsubsection{GES}
\label{sec:enc}
As mentioned, there is not yet a widely accepted way of determining exactly where the ex-situ stars originate from. Our sample of ex-situ stars can be compared directly to the sample of \Enc stars in \citet[][]{Massari2019}  (see their Figure 2) as they use the same potential as we do. All our stars are within the same region in \En vs \Lz space as their \Enc stars although some are very close to the edges. \citet[][]{Naidu2021} and \Montalban classify \Enc stars as those with $e>0.7$. This is the case for all of our ex-situ stars apart from \#8, 11 and 23. However in \autoref{fig:alpha_fe_enc}, we expect stars from the same galaxies to follow a similar trend but there is an almost $0.2$ dex difference in \alphafe between stars of the same [Fe/H] (stars \#3, 14 and \#1, 12). This could mean that the stars are not of the same origin. To be conservative we classify stars as \Enc only if they have a high \En and just around \Lz$=0$ as well as a similar trend in metallicity. These stars are marked by a green circle in \autoref{fig:Dynamics_enc} and \autoref{fig:alpha_fe_enc}. The stars that are classified as \Enc are green (have a green edge) in these two figures and stars that are not in both classifications are blue. This is also the colour coding in \autoref{fig:Ages}. We note that star \#22 is part of this selection but as discussed in \autoref{sec:peculiar} the age of this star is likely inaccurate and it is excluded from the mean age calculation. The mean age of our selected \Enc stars is \finalage.  

This might be a too conservative classification of \Enc stars. Especially the lower \En boundary in \autoref{fig:Dynamics_enc} could be excluding stars that appear to be \Enc stars in for example \autoref{fig:alpha_fe_enc}. 
However, the lower \En region is also where in-situ stars are potential contaminants and reducing the \En boundary could lead too a less pure \Enc sample. We therefore chose to be conservative in our selection and only include stars at high \En. 
As more studies on the merger galaxies are made it is possible that more stars from our sample turns out to be \Enc stars. 
We present the ages of individual stars making it easy for later works to use the ages of the stars should improved selection criteria be determined. 
For reference, we note that the mean age of the full sample (excluding \#22) is $10.1\pm0.7$ Gyr and the mean age of the stars with $e>0.7$ is $9.8\pm0.8$ Gyr.

\subsubsection*{Comparison with \Montalban}
\begin{figure}
	\centering
	\includegraphics[width=8.8cm]{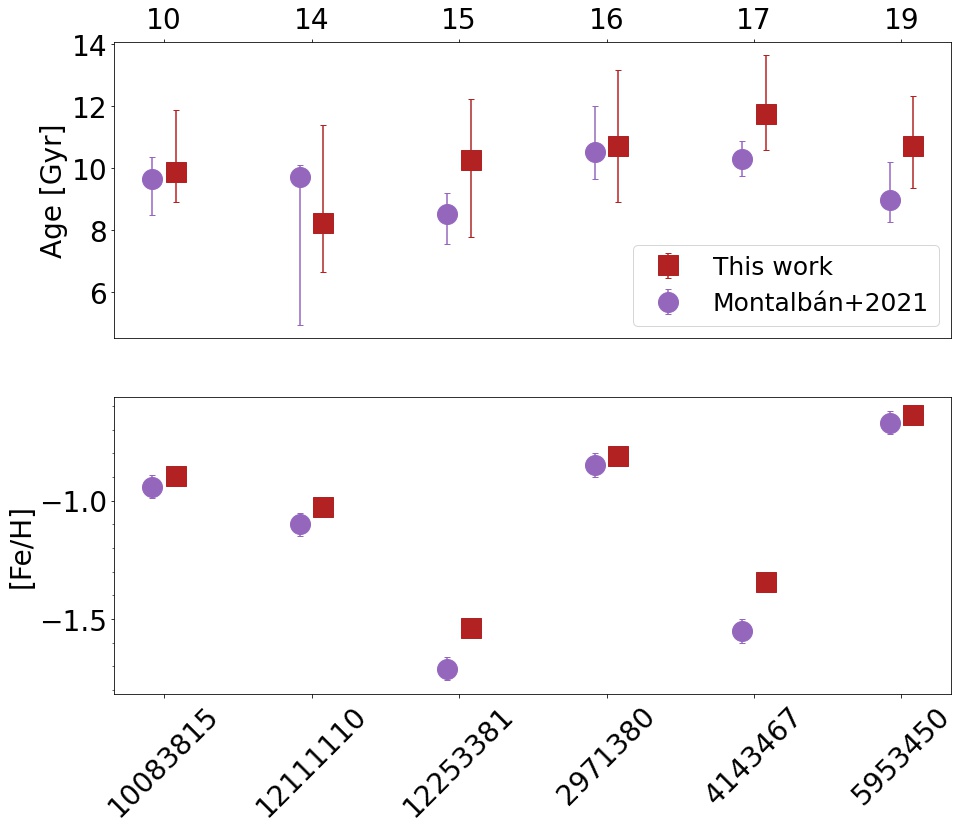}
	\caption{Comparison between our results and those of the same stars from \Montalban. Top: ages. Bottom: iron abundance. While \Montalban uses \APOGEE DR14, we use DR16. All stars are \kepler stars with KIC ID's on x-axis of bottom panel and indexes matching our figures on top for cross identification with the other figures presented in this work.}
	\label{fig:compare}
\end{figure}

\Montalban did a similar study to ours but used only RGB \kepler stars, different selections (see \autoref{tab:selection}), and a different code to calculate the ages. They further used \Gaia DR2 \citep{GaiaCollaboration2018} and \APOGEE DR14 \citep{Majewski2017} where we use eDR3 and DR16, respectively. In our ex-situ sample we find the same stars as they do except for KIC 8869235 because this was pruned in our selection due to a bad flag in \texttt{STAR$\_$WARN}. The comparison between our ages and the ages presented by \Montalban is shown in \autoref{fig:compare}. The ages agree well between the two studies and differences can be attributed to slightly different metallicities, and different software and models. Although we get similar ages for the stars we do not classify the same stars as \Enc but we do classify them as ex-situ. From the sample selected by \Montalban we do not classify star \#10, 17 and 19 as \Enc. On the other hand, as mentioned above, if we were to use the criteria proposed by \Montalban all stars except \#8, 11 and 23 would be \Enc stars. While our selection might be too conservative it is possible that the one presented in \Montalban is too broad.  
\Montalban calculated a mean age of the \Enc stars of $9.7\pm0.6$ Gyr and we get \finalage for our selection of \Enc stars. The age of our stars with $e>0.7$ is $9.8\pm0.8$ Gyr. All these ages are in agreement further enhancing the conclusion that the merger time is less than ${\sim}10$ Gyr ago.

\subsubsection{Kraken}
\label{sec:kraken}
Another galaxy some of the stars could belong to is Kraken \citep[][]{Kruijssen2019,Massari2019,Kruijssen2020}. Kraken is expected to have had similar mass as \Enc (${\sim}3\cdot10^8$M$_\odot$) and to have merged with the nascent Milky Way before \Enc. 
These stars are located at low \En and at $\lesssim7$ kpc from the galactic centre. Star \#8 is the closest star in our sample to the Galactic centre at ${\sim}4$ kpc. It is, however, at higher \En than we expect the Kraken stars to be based on the globular clusters from \citet[][]{Massari2019}. It might, therefore, originate from somewhere else. Stars \#3, 4, 7, and possibly 12 are closer to the expected \En for Kraken and stars \#4 and 7 are furthermore closer than ${\sim}7$ kpc to the Galactic centre. It is currently not clear where the boundaries of the Kraken remnants are and we can therefore not classify any of our stars as such with certainty. The accretion time of the Kraken was estimated to be $10.9^{+0.4}_{-0.7}$ Gyr by \citet[][]{Kruijssen2020} which agrees well with all these stars being old according to our calculations. Until we have more stars in this region or a tighter constraint on their properties compared to other ex-situ stars, it is unknown if and how many Kraken stars there are in our sample. If we assume stars \#3, 4, 7, and 12 are Kraken stars the mean age is $9.2_{-1.7}^{+1.9}$ Gyr, which is well within the estimate made by \citet[][]{Kruijssen2020}.

\subsubsection{Other possible origins}
\label{sec:other}
The stars that are not \Enc stars (or Kraken stars) could have different origins. There are some that are close to our selection criteria boundaries in both \alphafe and [Al/Fe] (see stars \#1, 11, 12, and 18 in \autoref{fig:alpha} and \autoref{fig:mgmn}), which could make them in-situ stars. It is also possible that they belong to Thamnos \citep[][]{Koppelman2019} or Sequoia \citep[][]{Myeong2019} or some other yet undefined merger event. 
However, there is no clear characterization yet that would allow us to do a firm statistically based membership probability or association and it is not possible to confidently classify their origin.
As more observations comes in from surveys such as \Gaia and \APOGEE it might be possible to classify them in the future. With more stars becoming available with \TESS and other asteroseismic surveys we will also be able to get ages for more of these ex-situ stars, which might make it possible to make classifications based on the ages of the stars in the future. For now we present the ages of the stars and leave further classification for future works.

\section{Conclusion}
\label{sec:conclusion} 
In this paper, we examine a sample of nearly $12,000$ \kepler and \kt stars and we find $23$ stars of ex-situ origin. Our selection is based on angular momentum (\Lz), energy (\En), and the \alphafe, [Fe/H], [Al/Fe], and [Mg/Mn] abundances making it very stringent to insure a high likelihood of choosing only ex-situ stars.

Using asteroseismology from \kepler and \kt, 6D phase space data from \Gaia eDR3, along with temperature and metallicity from \APOGEE DR16, we derive their stellar properties using the Bayesian framework of \BASTA. We present ages for all $23$ ex-situ stars as well as $55$ in-situ stars.

All the ex-situ stars are consistent with being older than $8$ Gyr (except for one that is likely a young $\alpha$-rich star), which is expected because most massive mergers took place more than $8$ Gyr ago.   

We make a conservative classification of the ex-situ stars and denote eight of them to be \Enc stars with a mean age of \finalage, which agrees with current estimates of a merger time around $8-10$ Gyr ago. We compare our stellar ages to those found be \citet{Montalban2021} and demonstrate similar results but based on different methods. The classification of \Enc stars between the studies deviate but we get similar mean ages of the \Enc stars regardless.  

The remaining $15$ stars are ex-situ stars of debatable origin and some of them could also belong to \Enc. It is possible that four of these stars belong to the dwarf galaxy Kraken and their mean ages are estimated to be $9.2_{-1.7}^{+1.9}$ Gyr in agreement with estimates from \citet[][]{Kruijssen2020}.

With more data becoming available from for example the \TESS mission, it might be possible to find even more asteroseismic merger remnant stars in the future. With more stars with precise ages it might be possible to distinguish stars from different origins based on not only their kinematics and chemistry but also their ages.

Due to the large impact mergers have had on the Milky Way's evolution, the stars presented in this work contribute important information to better understand how the mergers happened and what impacted they have had on our present day Milky Way.

\section*{Acknowledgements}
The authors thank Joel Zinn for kindly providing the K2 GAP DR3 data prior to its public release. The authors thanks Eduardo Balbinot for providing E and Lz values for the sample of Gaia eDR3 stars.

Funding for the Stellar Astrophysics Centre is provided by The Danish National Research Foundation (Grant agreement no.: DNRF106).
AH acknowledges support from a Spinoza prize from the Netherlands Research Council (NWO).
HHK gratefully acknowledges financial support from a Fellowship at the Institute for Advanced Study.
AS acknowledge support from the European Research Council Consolidator Grant funding scheme (project ASTEROCHRONOMETRY, G.A. n. 772293, \url{http://www.asterochronometry.eu}).
JMDK gratefully acknowledges funding from the Deutsche Forschungsgemeinschaft (DFG, German Research Foundation) through an Emmy Noether Research Group (grant number KR4801/1-1), as well as from the European Research Council (ERC) under the European Union's Horizon 2020 research and innovation programme via the ERC Starting Grant MUSTANG (grant agreement number 714907). 
JY acknowledges partial support from ERC Synergy Grant WHOLE SUN 810218. 

This work has made use of data from the European Space Agency (ESA) mission
{\it Gaia} (\url{https://www.cosmos.esa.int/gaia}), processed by the {\it Gaia}
Data Processing and Analysis Consortium (DPAC,
\url{https://www.cosmos.esa.int/web/gaia/dpac/consortium}). Funding for the DPAC
has been provided by national institutions, in particular the institutions
participating in the {\it Gaia} Multilateral Agreement.
This publication makes use of data products from the Two Micron All Sky Survey, which is a joint project of the University of Massachusetts and the Infrared Processing and Analysis Center/California Institute of Technology, funded by the National Aeronautics and Space Administration and the National Science Foundation.

\section*{Software}
The research for this publication was coded in \textsc{python} \citep[v. 3.8.5;][]{Python}
and included its packages \textsc{astropy} \citep[v. 4.0.2;][]{astropy:2013,astropy:2018}, \textsc{galpy} \citep[v. 1.6.0;][]{Bovy2015}
\textsc{matplotlib} \citep[v. 3.3.2;][]{matplotlib2007}, \textsc{numpy} \citep[v. 1.19.2;][]{numpy}, \textsc{vaex} \citep[v. 3.0.0;][]{vaex}, \textsc{IPython} \citep[v. 7.19.0;][]{ipython}, \textsc{Jupyter} \citep[v. 2.2.6;][]{jupyter} and \textsc{Spyder} \citep[v. 4.1.5;][]{spyder}. For frequency peak bagging we used \textsc{PBjam} \citep[][]{PBJam} and for stellar properties determination we used \BASTA \citep{Aguirre2021}.

\section*{Data Availability}


The data underlying this article will be available in the article and in its online supplementary material upon publication.



\bibliographystyle{mnras}
\bibliography{Library} 




\appendix

\section{Stars in our sample }
\begin{onecolumn}

\begin{figure}
	\centering
	\includegraphics[angle=-90,width=18cm]{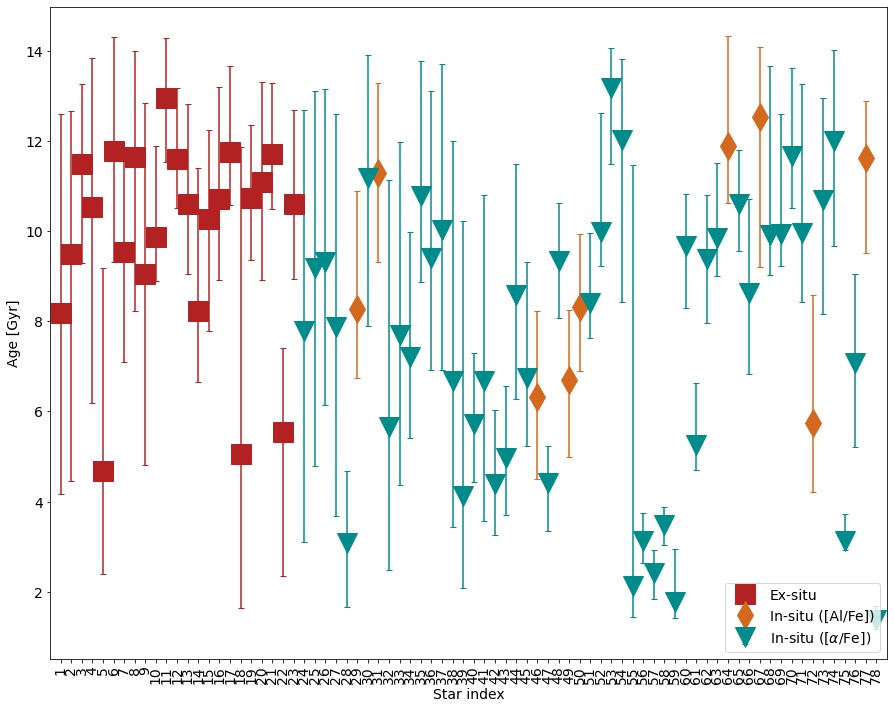}
	\caption{Ages of the stars with uncertainties. Indices match the ones used in other figures. Symbols are explained in \autoref{sec:selection}.} 
	\label{fig:Agesx3}
\end{figure}

\LTcapwidth=\textwidth
\begin{landscape}
\renewcommand{\arraystretch}{1.5}

\begin{longtable}{ccccccrrrcc}

\caption{Table of stars from our sample. The columns contain the index of the stars used throughout the paper (INDEX), the ID of the stars where "Kepler" denotes stars with KIC ID's and K2Cx denotes stars from the \kt mission with x corresponding to the campaign number and corresponding EPIC ID.  The asteroseismic parameters from PBJam or the SYD pipeline in $\mu$Hz. The following column is the distance to the Galactic centre in kpc and the eccentricity of the stars. The distance, radius and ages with uncertainties (16th and 84th percentile from the posterior distribution) are derived from \BASTA in kpc, R$_\odot$ and Gyr, respectively. The second to last column contain information about how each star was fitted. "Indi" denotes stars that had individual frequencies available, "Seis" indicates stars that only had \numax and \dnu available and "(no) dist" denotes stars where the distance was (not) fitted. The last column contains our classification of the star. } \label{tab:All stars}\\
%
INDEX & ID &  \dnu  & \numax  &  R$_{\text{GC}}$ &   e  & Distance & R$_{*}$  & Age  & FIT & Classification \\
\hline
1	&	K2C1 201609020	&	$	3.99	\pm	0.02	$&$	31.63	\pm	0.91	$	&	$	8.35	\pm	0.01	$	&	$	0.78	\pm	0.02	$	&	$	2910	_{-	64	}^{+	91	}$	&	$	10.1	_{-	0.4	}^{+	0.7	}$	&	$	8.19	_{-	4.02	}^{+	4.41	}$	&		Indi,	no	dist	&	Ex-situ	\\
2	&	K2C1 201709805	&	$	3.68	\pm	0.03	$&$	30.06	\pm	0.86	$	&	$	8.49	\pm	0.01	$	&	$	0.97	\pm	0.01	$	&	$	3126	_{-	54	}^{+	68	}$	&	$	10.4	_{-	0.3	}^{+	0.8	}$	&	$	9.49	_{-	5.03	}^{+	3.17	}$	&		Indi,	no	dist	&	\Enc	\\
3	&	K2C3 205934106	&	$	13.24	\pm	0.04	$&$	145.94	\pm	4.28	$	&	$	8.02	\pm	0.01	$	&	$	0.77	\pm	0.01	$	&	$	465	_{-	7	}^{+	7	}$	&	$	4.4	_{-	0.1	}^{+	0.1	}$	&	$	11.48	_{-	2.19	}^{+	1.78	}$	&		Indi,	dist	&	Ex-situ		\\
4	&	K2C3 206077646	&	$	0.64	\pm	0.12	$&$	4.79	\pm	0.62	$	&	$	6.42	\pm	0.15	$	&	$	0.92	\pm	0.07	$	&	$	5134	_{-	172	}^{+	183	}$	&	$	19.5	_{-	1.4	}^{+	1.6	}$	&	$	10.54	_{-	4.35	}^{+	5.49	}$	&		No seis,	dist	&	Ex-situ		\\
5	&	K2C3 206516677	&	$	3.96	\pm	0.06	$&$	30.19	\pm	2.05	$	&	$	7.95	\pm	0.01	$	&	$	0.99	\pm	0.01	$	&	$	957	_{-	40	}^{+	42	}$	&	$	11.0	_{-	0.8	}^{+	0.9	}$	&	$	4.67	_{-	2.27	}^{+	4.50	}$	&		Seis,	no	dist	&	\Enc	\\
6	&	K2C4 211001845	&	$	3.49	\pm	0.08	$&$	27.59	\pm	0.98	$	&	$	10.23	\pm	0.08	$	&	$	0.98	\pm	0.03	$	&	$	2289	_{-	38	}^{+	44	}$	&	$	10.5	_{-	0.2	}^{+	0.3	}$	&	$	11.78	_{-	2.47	}^{+	4.06	}$	&	\footnote{K2 GAP seismology}	Seis,	no	dist	&	\Enc	\\
7	&	K2C6 212319585	&	$	3.73	\pm	0.02	$&$	27.00	\pm	1.28	$	&	$	7.00	\pm	0.05	$	&	$	0.72	\pm	0.03	$	&	$	2302	_{-	30	}^{+	35	}$	&	$	10.3	_{-	0.2	}^{+	0.3	}$	&	$	9.53	_{-	2.43	}^{+	2.31	}$	&		Indi,	dist	&	Ex-situ		\\
8	&	K2C7 213616808	&	$	1.23	\pm	0.05	$&$	6.32	\pm	1.96	$	&	$	3.98	\pm	0.62	$	&	$	0.44	\pm	0.16	$	&	$	4810	_{-	93	}^{+	98	}$	&	$	21.7	_{-	0.7	}^{+	0.8	}$	&	$	11.65	_{-	3.41	}^{+	2.36	}$	&		Seis,	dist	&	Ex-situ		\\
9	&	K2C8 220238966	&	$	4.10	\pm	0.03	$&$	30.78	\pm	2.09	$	&	$	9.02	\pm	0.04	$	&	$	0.93	\pm	0.03	$	&	$	2686	_{-	53	}^{+	70	}$	&	$	9.7	_{-	0.3	}^{+	0.6	}$	&	$	9.05	_{-	4.23	}^{+	3.78	}$	&		Indi,	no	dist	&	\Enc	\\
10	&	Kepler 10083815	&	$	2.59	\pm	0.01	$&$	17.99	\pm	0.38	$	&	$	8.14	\pm	0.01	$	&	$	0.93	\pm	0.01	$	&	$	2415	_{-	21	}^{+	20	}$	&	$	13.0	_{-	0.2	}^{+	0.2	}$	&	$	9.87	_{-	0.97	}^{+	2.00	}$	&		Indi,	dist	&	Ex-situ		\\
11	&	Kepler 10460723	&	$	3.14	\pm	0.02	$&$	22.97	\pm	0.45	$	&	$	8.07	\pm	0.01	$	&	$	0.49	\pm	0.01	$	&	$	2400	_{-	19	}^{+	16	}$	&	$	11.1	_{-	0.1	}^{+	0.1	}$	&	$	12.94	_{-	1.41	}^{+	1.34	}$	&	\footnote{\label{note1}Evolutionary phase changed to RGB}	Indi,	dist	&	Ex-situ		\\
12	&	Kepler 11563791	&	$	5.04	\pm	0.03	$&$	43.03	\pm	0.51	$	&	$	8.13	\pm	0.01	$	&	$	0.73	\pm	0.01	$	&	$	978	_{-	7	}^{+	6	}$	&	$	8.2	_{-	0.1	}^{+	0.1	}$	&	$	11.60	_{-	1.09	}^{+	1.57	}$	&		Indi,	dist	&	Ex-situ		\\
13	&	Kepler 11566038	&	$	3.95	\pm	0.03	$&$	31.36	\pm	0.32	$	&	$	8.18	\pm	0.01	$	&	$	0.79	\pm	0.01	$	&	$	1945	_{-	20	}^{+	22	}$	&	$	9.7	_{-	0.2	}^{+	0.2	}$	&	$	10.60	_{-	1.56	}^{+	2.21	}$	&		Indi,	dist	&	Ex-situ		\\
14	&	Kepler 12111110	&	$	3.75	\pm	0.02	$&$	29.57	\pm	0.37	$	&	$	8.34	\pm	0.03	$	&	$	0.99	\pm	0.01	$	&	$	3052	_{-	70	}^{+	53	}$	&	$	10.5	_{-	0.4	}^{+	0.3	}$	&	$	8.22	_{-	1.57	}^{+	3.18	}$	&		Indi,	dist	&	\Enc		\\
15	&	Kepler 12253381	&	$	3.03	\pm	0.02	$&$	22.00	\pm	0.40	$	&	$	8.24	\pm	0.01	$	&	$	0.97	\pm	0.01	$	&	$	2795	_{-	42	}^{+	41	}$	&	$	11.6	_{-	0.2	}^{+	0.3	}$	&	$	10.27	_{-	2.48	}^{+	1.96	}$	&		Indi,	no	dist	&	\Enc	\\
16	&	Kepler 2971380	&	$	7.88	\pm	0.03	$&$	77.26	\pm	0.70	$	&	$	7.75	\pm	0.01	$	&	$	0.91	\pm	0.01	$	&	$	1552	_{-	19	}^{+	25	}$	&	$	6.2	_{-	0.1	}^{+	0.1	}$	&	$	10.72	_{-	1.81	}^{+	2.46	}$	&		Indi,	no	dist	&	\Enc	\\
17	&	Kepler 4143467	&	$	5.63	\pm	0.02	$&$	48.96	\pm	0.57	$	&	$	7.77	\pm	0.01	$	&	$	0.93	\pm	0.01	$	&	$	2009	_{-	18	}^{+	21	}$	&	$	7.6	_{-	0.1	}^{+	0.1	}$	&	$	11.75	_{-	1.17	}^{+	1.91	}$	&		Indi,	dist	&	Ex-situ		\\
18	&	Kepler 5446927	&	$	2.91	\pm	0.03	$&$	21.88	\pm	0.38	$	&	$	7.85	\pm	0.01	$	&	$	0.88	\pm	0.01	$	&	$	2184	_{-	107	}^{+	379	}$	&	$	13.2	_{-	1.3	}^{+	1.8	}$	&	$	5.06	_{-	3.41	}^{+	6.80	}$	&		Indi,	no	dist	&	Ex-situ	\\
19	&	Kepler 5953450	&	$	12.69	\pm	0.05	$&$	140.87	\pm	0.83	$	&	$	7.94	\pm	0.01	$	&	$	0.98	\pm	0.01	$	&	$	1250	_{-	6	}^{+	9	}$	&	$	4.6	_{-	0.1	}^{+	0.1	}$	&	$	10.73	_{-	1.36	}^{+	1.62	}$	&		Indi,	dist	&	Ex-situ		\\
20	&	Kepler 6865157	&	$	1.57	\pm	0.07	$&$	9.05	\pm	0.28	$	&	$	7.99	\pm	0.01	$	&	$	0.75	\pm	0.01	$	&	$	3177	_{-	41	}^{+	46	}$	&	$	18.4	_{-	0.4	}^{+	0.5	}$	&	$	11.08	_{-	2.17	}^{+	2.21	}$	&		Seis,	dist	&	Ex-situ		\\
21	&	Kepler 7948268	&	$	11.45	\pm	0.04	$&$	120.45	\pm	0.82	$	&	$	7.97	\pm	0.01	$	&	$	0.80	\pm	0.01	$	&	$	1283	_{-	12	}^{+	10	}$	&	$	4.8	_{-	0.1	}^{+	0.1	}$	&	$	11.71	_{-	1.22	}^{+	1.57	}$	&		Indi,	dist	&	Ex-situ		\\
22	&	Kepler 8694070	&	$	4.63	\pm	0.04	$&$	33.66	\pm	0.42	$	&	$	8.04	\pm	0.01	$	&	$	0.88	\pm	0.01	$	&	$	2701	_{-	41	}^{+	41	}$	&	$	9.7	_{-	0.3	}^{+	1.1	}$	&	$	5.54	_{-	3.19	}^{+	1.88	}$	&		Indi,	no	dist	&	\Enc	\\
23	&	Kepler 9339711	&	$	2.84	\pm	0.01	$&$	20.51	\pm	0.31	$	&	$	8.03	\pm	0.01	$	&	$	0.45	\pm	0.01	$	&	$	2266	_{-	26	}^{+	21	}$	&	$	12.1	_{-	0.2	}^{+	0.2	}$	&	$	10.61	_{-	1.68	}^{+	2.08	}$	&		Indi,	dist	&	Ex-situ		\\
24	&	K2C11 204785972	&	$	3.34	\pm	0.17	$&$	23.10	\pm	1.69	$	&	$	6.55	\pm	0.04	$	&	$	0.68	\pm	0.01	$	&	$	2000	_{-	59	}^{+	69	}$	&	$	12.0	_{-	0.8	}^{+	1.7	}$	&	$	7.79	_{-	4.69	}^{+	4.89	}$	&		Seis,	no	dist	&	In-situ	\\
25	&	K2C17 251512185	&	$	4.73	\pm	0.34	$&$	40.18	\pm	0.77	$	&	$	7.73	\pm	0.01	$	&	$	0.85	\pm	0.03	$	&	$	1554	_{-	42	}^{+	51	}$	&	$	9.1	_{-	0.5	}^{+	0.8	}$	&	$	9.18	_{-	4.38	}^{+	3.92	}$	&		Seis,	no	dist	&	In-situ	\\
26	&	K2C2 204298932	&	$	1.69	\pm	0.05	$&$	10.30	\pm	0.30	$	&	$	6.65	\pm	0.04	$	&	$	0.97	\pm	0.03	$	&	$	1782	_{-	69	}^{+	76	}$	&	$	18.3	_{-	0.7	}^{+	1.0	}$	&	$	9.31	_{-	3.16	}^{+	3.84	}$	&		Seis,	no	dist	&	In-situ	\\
27	&	K2C2 204466120	&	$	1.68	\pm	0.05	$&$	10.97	\pm	1.59	$	&	$	4.90	\pm	0.25	$	&	$	0.47	\pm	0.03	$	&	$	4045	_{-	116	}^{+	135	}$	&	$	18.7	_{-	0.9	}^{+	1.4	}$	&	$	7.87	_{-	4.18	}^{+	4.73	}$	&		Seis,	no	dist	&	In-situ	\\
28	&	K2C2 204966489	&	$	2.26	\pm	0.11	$&$	18.45	\pm	0.37	$	&	$	4.42	\pm	0.24	$	&	$	0.27	\pm	0.02	$	&	$	3212	_{-	100	}^{+	90	}$	&	$	16.7	_{-	1.0	}^{+	1.5	}$	&	$	3.08	_{-	1.42	}^{+	1.59	}$	&		Seis,	no	dist	&	In-situ	\\
29	&	K2C2 205083494	&	$	1.21	\pm	0.03	$&$	6.80	\pm	0.20	$	&	$	5.26	\pm	0.14	$	&	$	0.40	\pm	0.02	$	&	$	3225	_{-	53	}^{+	51	}$	&	$	22.3	_{-	0.6	}^{+	0.6	}$	&	$	8.27	_{-	1.52	}^{+	2.61	}$	&		Seis,	dist	&	In-situ		\\
30	&	K2C3 205972576	&	$	4.08	\pm	0.12	$&$	28.51	\pm	6.21	$	&	$	7.17	\pm	0.05	$	&	$	0.82	\pm	0.03	$	&	$	2697	_{-	43	}^{+	49	}$	&	$	10.2	_{-	0.4	}^{+	0.3	}$	&	$	11.18	_{-	3.29	}^{+	2.71	}$	&		Seis,	dist	&	In-situ		\\
31	&	K2C3 205997746	&	$	5.69	\pm	0.03	$&$	51.06	\pm	0.86	$	&	$	7.59	\pm	0.02	$	&	$	0.73	\pm	0.02	$	&	$	1859	_{-	21	}^{+	22	}$	&	$	7.6	_{-	0.1	}^{+	0.2	}$	&	$	11.29	_{-	1.98	}^{+	1.98	}$	&		Indi,	dist	&	In-situ		\\
32	&	K2C3 206011766	&	$	1.22	\pm	0.09	$&$	7.34	\pm	0.20	$	&	$	6.66	\pm	0.09	$	&	$	0.63	\pm	0.04	$	&	$	4145	_{-	168	}^{+	351	}$	&	$	23.2	_{-	2.1	}^{+	3.0	}$	&	$	5.66	_{-	3.17	}^{+	5.47	}$	&		Seis,	no	dist	&	In-situ	\\
33	&	K2C6 212297999	&	$	2.60	\pm	0.07	$&$	19.31	\pm	0.34	$	&	$	7.00	\pm	0.04	$	&	$	0.74	\pm	0.02	$	&	$	2485	_{-	73	}^{+	80	}$	&	$	13.8	_{-	0.7	}^{+	1.0	}$	&	$	7.69	_{-	3.32	}^{+	4.28	}$	&		Seis,	no	dist	&	In-situ	\\
34	&	K2C6 212302713	&	$	3.94	\pm	0.02	$&$	31.94	\pm	0.79	$	&	$	6.62	\pm	0.08	$	&	$	0.56	\pm	0.04	$	&	$	3175	_{-	48	}^{+	46	}$	&	$	10.6	_{-	0.3	}^{+	0.3	}$	&	$	7.21	_{-	1.80	}^{+	2.76	}$	&		Indi,	dist	&	In-situ		\\
35	&	K2C7 213463719	&	$	5.80	\pm	0.02	$&$	54.92	\pm	1.38	$	&	$	6.54	\pm	0.07	$	&	$	0.69	\pm	0.03	$	&	$	1825	_{-	21	}^{+	23	}$	&	$	7.9	_{-	0.2	}^{+	0.2	}$	&	$	10.78	_{-	1.92	}^{+	2.98	}$	&		Indi,	dist	&	In-situ		\\
36	&	K2C7 213523425	&	$	4.74	\pm	0.04	$&$	43.10	\pm	1.57	$	&	$	5.90	\pm	0.09	$	&	$	0.64	\pm	0.03	$	&	$	2480	_{-	48	}^{+	52	}$	&	$	9.2	_{-	0.3	}^{+	0.3	}$	&	$	9.40	_{-	2.48	}^{+	3.70	}$	&		Indi,	dist	&	In-situ		\\
37	&	K2C7 213532050	&	$	8.80	\pm	0.04	$&$	88.85	\pm	4.22	$	&	$	7.20	\pm	0.02	$	&	$	0.61	\pm	0.02	$	&	$	1166	_{-	21	}^{+	28	}$	&	$	5.9	_{-	0.2	}^{+	0.2	}$	&	$	10.01	_{-	3.10	}^{+	3.69	}$	&		Indi,	no	dist	&	In-situ	\\
38	&	K2C7 213632986	&	$	1.29	\pm	0.04	$&$	7.89	\pm	0.27	$	&	$	5.35	\pm	0.13	$	&	$	0.69	\pm	0.01	$	&	$	3942	_{-	141	}^{+	228	}$	&	$	22.4	_{-	1.5	}^{+	2.1	}$	&	$	6.67	_{-	3.22	}^{+	5.32	}$	&		Seis,	no	dist	&	In-situ	\\
39	&	K2C7 213651916	&	$	3.34	\pm	2.81	$&$	20.35	\pm	2.62	$	&	$	5.60	\pm	0.11	$	&	$	0.45	\pm	0.02	$	&	$	3960	_{-	181	}^{+	160	}$	&	$	14.6	_{-	2.2	}^{+	1.4	}$	&	$	4.13	_{-	2.03	}^{+	6.09	}$	&		Seis,	no	dist	&	In-situ	\\
40	&	K2C7 213764390	&	$	1.38	\pm	0.15	$&$	9.16	\pm	0.38	$	&	$	5.15	\pm	0.14	$	&	$	0.54	\pm	0.03	$	&	$	3310	_{-	56	}^{+	56	}$	&	$	21.4	_{-	0.7	}^{+	0.7	}$	&	$	5.71	_{-	1.27	}^{+	1.58	}$	&		Seis,	dist	&	In-situ		\\
41	&	K2C7 213840500	&	$	3.07	\pm	0.06	$&$	24.80	\pm	0.91	$	&	$	5.16	\pm	0.20	$	&	$	0.53	\pm	0.04	$	&	$	2765	_{-	86	}^{+	117	}$	&	$	12.8	_{-	0.7	}^{+	0.9	}$	&	$	6.66	_{-	3.08	}^{+	4.14	}$	&		Seis,	no	dist	&	In-situ	\\
42	&	K2C7 213853964	&	$	4.31	\pm	0.03	$&$	36.12	\pm	2.93	$	&	$	5.52	\pm	0.15	$	&	$	0.59	\pm	0.04	$	&	$	2897	_{-	51	}^{+	52	}$	&	$	10.2	_{-	0.3	}^{+	0.3	}$	&	$	4.40	_{-	1.14	}^{+	1.64	}$	&		Indi,	dist	&	In-situ		\\
43	&	K2C7 214605185	&	$	1.26	\pm	0.04	$&$	8.08	\pm	0.24	$	&	$	4.74	\pm	0.23	$	&	$	0.39	\pm	0.02	$	&	$	3670	_{-	71	}^{+	72	}$	&	$	23.8	_{-	0.9	}^{+	0.9	}$	&	$	4.97	_{-	1.27	}^{+	1.59	}$	&		Seis,	dist	&	In-situ		\\
44	&	K2C7 219534658	&	$	4.74	\pm	0.02	$&$	43.01	\pm	1.78	$	&	$	6.16	\pm	0.08	$	&	$	0.59	\pm	0.01	$	&	$	2283	_{-	36	}^{+	38	}$	&	$	9.2	_{-	0.2	}^{+	0.3	}$	&	$	8.58	_{-	2.30	}^{+	2.91	}$	&		Indi,	dist	&	In-situ		\\
45	&	K2C8 220269276	&	$	6.27	\pm	0.02	$&$	65.79	\pm	2.53	$	&	$	9.29	\pm	0.05	$	&	$	0.79	\pm	0.04	$	&	$	2605	_{-	43	}^{+	41	}$	&	$	7.8	_{-	0.2	}^{+	0.2	}$	&	$	6.74	_{-	1.51	}^{+	2.58	}$	&		Indi,	dist	&	In-situ		\\
46	&	K2C8 220387868	&	$	3.84	\pm	0.04	$&$	32.20	\pm	1.01	$	&	$	8.75	\pm	0.02	$	&	$	0.72	\pm	0.03	$	&	$	1638	_{-	29	}^{+	34	}$	&	$	10.8	_{-	0.3	}^{+	0.4	}$	&	$	6.32	_{-	1.83	}^{+	1.90	}$	&		Indi,	dist	&	In-situ		\\
47	&	Kepler 10096113	&	$	4.22	\pm	0.03	$&$	36.31	\pm	0.59	$	&	$	8.37	\pm	0.03	$	&	$	0.93	\pm	0.01	$	&	$	2834	_{-	38	}^{+	41	}$	&	$	10.7	_{-	0.2	}^{+	0.3	}$	&	$	4.42	_{-	1.06	}^{+	0.83	}$	&		Indi,	no	dist	&	In-situ	\\
48	&	Kepler 10207078	&	$	4.57	\pm	0.02	$&$	40.30	\pm	0.65	$	&	$	8.06	\pm	0.01	$	&	$	1.00	\pm	0.01	$	&	$	1508	_{-	10	}^{+	9	}$	&	$	9.3	_{-	0.1	}^{+	0.1	}$	&	$	9.34	_{-	1.28	}^{+	1.27	}$	&		Indi,	dist	&	In-situ		\\
49	&	Kepler 10319045	&	$	1.18	\pm	0.02	$&$	6.53	\pm	0.24	$	&	$	8.06	\pm	0.01	$	&	$	0.63	\pm	0.01	$	&	$	3108	_{-	45	}^{+	51	}$	&	$	23.4	_{-	0.6	}^{+	0.7	}$	&	$	6.69	_{-	1.70	}^{+	1.56	}$	&		Seis,	dist	&	In-situ		\\
50	&	Kepler 10398120	&	$	1.59	\pm	0.10	$&$	8.58	\pm	0.27	$	&	$	8.05	\pm	0.01	$	&	$	0.84	\pm	0.01	$	&	$	1854	_{-	26	}^{+	27	}$	&	$	19.4	_{-	0.5	}^{+	0.5	}$	&	$	8.32	_{-	1.43	}^{+	1.62	}$	&		Seis,	dist	&	In-situ		\\
51	&	Kepler 10992126	&	$	1.33	\pm	0.14	$&$	7.83	\pm	0.22	$	&	$	8.16	\pm	0.01	$	&	$	0.97	\pm	0.01	$	&	$	1698	_{-	24	}^{+	24	}$	&	$	21.3	_{-	0.5	}^{+	0.5	}$	&	$	8.40	_{-	0.77	}^{+	1.56	}$	&		Seis,	dist	&	In-situ		\\
52	&	Kepler 11037292	&	$	2.40	\pm	0.01	$&$	17.01	\pm	0.25	$	&	$	8.16	\pm	0.01	$	&	$	0.65	\pm	0.01	$	&	$	2127	_{-	21	}^{+	23	}$	&	$	13.8	_{-	0.2	}^{+	0.2	}$	&	$	9.97	_{-	0.74	}^{+	2.66	}$	&		Indi,	dist	&	In-situ		\\
53	&	Kepler 11774651	&	$	4.60	\pm	0.02	$&$	39.78	\pm	0.43	$	&	$	8.29	\pm	0.01	$	&	$	0.72	\pm	0.01	$	&	$	2462	_{-	18	}^{+	18	}$	&	$	9.0	_{-	0.1	}^{+	0.1	}$	&	$	13.18	_{-	1.69	}^{+	0.89	}$	&		Indi,	dist	&	In-situ		\\
54	&	Kepler 12109442	&	$	4.11	\pm	0.13	$&$	28.57	\pm	0.83	$	&	$	8.13	\pm	0.01	$	&	$	0.98	\pm	0.01	$	&	$	1065	_{-	16	}^{+	16	}$	&	$	10.1	_{-	0.3	}^{+	0.3	}$	&	$	12.02	_{-	3.59	}^{+	1.80	}$	&		Seis,	no	dist	&	In-situ	\\
55	&	Kepler 12506245	&	$	2.97	\pm	0.02	$&$	19.17	\pm	0.69	$	&	$	8.26	\pm	0.01	$	&	$	0.64	\pm	0.01	$	&	$	3442	_{-	92	}^{+	68	}$	&	$	14.8	_{-	2.6	}^{+	0.7	}$	&	$	2.14	_{-	0.70	}^{+	9.32	}$	&		Indi,	no	dist	&	In-situ	\\
56	&	Kepler 1726211	&	$	3.76	\pm	0.03	$&$	30.70	\pm	0.63	$	&	$	7.87	\pm	0.01	$	&	$	0.70	\pm	0.01	$	&	$	1308	_{-	15	}^{+	17	}$	&	$	11.9	_{-	0.3	}^{+	0.3	}$	&	$	3.14	_{-	0.49	}^{+	0.61	}$	&		Indi,	dist	&	In-situ		\\
57	&	Kepler 2165615	&	$	4.15	\pm	0.02	$&$	37.78	\pm	0.98	$	&	$	7.75	\pm	0.01	$	&	$	0.59	\pm	0.02	$	&	$	3175	_{-	40	}^{+	52	}$	&	$	11.4	_{-	0.2	}^{+	0.3	}$	&	$	2.43	_{-	0.59	}^{+	0.50	}$	&		Indi,	dist	&	In-situ		\\
58	&	Kepler 2301577	&	$	4.27	\pm	0.03	$&$	35.98	\pm	0.95	$	&	$	7.74	\pm	0.01	$	&	$	0.99	\pm	0.01	$	&	$	3248	_{-	27	}^{+	26	}$	&	$	10.9	_{-	0.2	}^{+	0.2	}$	&	$	3.48	_{-	0.45	}^{+	0.41	}$	&		Indi,	dist	&	In-situ		\\
59	&	Kepler 2444790	&	$	2.90	\pm	0.02	$&$	18.91	\pm	0.80	$	&	$	7.78	\pm	0.02	$	&	$	0.62	\pm	0.01	$	&	$	4281	_{-	80	}^{+	73	}$	&	$	15.6	_{-	0.8	}^{+	0.4	}$	&	$	1.78	_{-	0.36	}^{+	1.16	}$	&		Indi,	no	dist	&	In-situ	\\
60	&	Kepler 2571323	&	$	4.65	\pm	0.02	$&$	39.98	\pm	0.30	$	&	$	7.72	\pm	0.01	$	&	$	0.74	\pm	0.01	$	&	$	2630	_{-	21	}^{+	23	}$	&	$	9.0	_{-	0.1	}^{+	0.1	}$	&	$	9.67	_{-	1.38	}^{+	1.16	}$	&		Indi,	dist	&	In-situ		\\
61	&	Kepler 2714397	&	$	4.19	\pm	0.03	$&$	33.08	\pm	0.64	$	&	$	7.95	\pm	0.01	$	&	$	0.80	\pm	0.01	$	&	$	967	_{-	7	}^{+	7	}$	&	$	10.5	_{-	0.2	}^{+	0.1	}$	&	$	5.25	_{-	0.56	}^{+	1.39	}$	&		Indi,	dist	&	In-situ		\\
62	&	Kepler 2831815	&	$	5.61	\pm	0.03	$&$	52.88	\pm	0.50	$	&	$	7.73	\pm	0.01	$	&	$	0.70	\pm	0.01	$	&	$	1867	_{-	15	}^{+	18	}$	&	$	8.1	_{-	0.1	}^{+	0.2	}$	&	$	9.38	_{-	1.41	}^{+	1.42	}$	&		Indi,	dist	&	In-situ		\\
63	&	Kepler 5371173	&	$	5.09	\pm	0.02	$&$	45.78	\pm	0.32	$	&	$	7.91	\pm	0.01	$	&	$	0.66	\pm	0.01	$	&	$	1669	_{-	9	}^{+	11	}$	&	$	8.5	_{-	0.1	}^{+	0.1	}$	&	$	9.84	_{-	0.84	}^{+	1.67	}$	&		Indi,	dist	&	In-situ		\\
64	&	Kepler 5698156	&	$	1.68	\pm	0.03	$&$	9.73	\pm	0.31	$	&	$	7.92	\pm	0.01	$	&	$	0.33	\pm	0.01	$	&	$	1352	_{-	13	}^{+	14	}$	&	$	17.2	_{-	0.3	}^{+	0.3	}$	&	$	11.89	_{-	1.28	}^{+	2.42	}$	&		Seis,	dist	&	In-situ		\\
65	&	Kepler 5792889	&	$	13.15	\pm	0.05	$&$	150.71	\pm	0.88	$	&	$	7.95	\pm	0.01	$	&	$	0.62	\pm	0.01	$	&	$	1247	_{-	12	}^{+	10	}$	&	$	4.5	_{-	0.1	}^{+	0.1	}$	&	$	10.60	_{-	1.05	}^{+	1.19	}$	&		Indi,	dist	&	In-situ		\\
66	&	Kepler 6267115	&	$	2.03	\pm	0.01	$&$	13.52	\pm	0.25	$	&	$	7.80	\pm	0.01	$	&	$	0.83	\pm	0.01	$	&	$	2774	_{-	31	}^{+	35	}$	&	$	16.2	_{-	0.3	}^{+	0.3	}$	&	$	8.63	_{-	1.80	}^{+	2.08	}$	&	\textsuperscript{\ref{note1}}	Indi,	dist	&	In-situ		\\
67	&	Kepler 7191496	&	$	2.47	\pm	0.02	$&$	16.23	\pm	0.24	$	&	$	7.90	\pm	0.01	$	&	$	0.71	\pm	0.01	$	&	$	2308	_{-	35	}^{+	40	}$	&	$	13.3	_{-	0.2	}^{+	0.4	}$	&	$	12.53	_{-	3.33	}^{+	1.55	}$	&		Indi,	no	dist	&	In-situ	\\
68	&	Kepler 7502070	&	$	4.00	\pm	0.02	$&$	33.77	\pm	0.68	$	&	$	7.85	\pm	0.01	$	&	$	0.67	\pm	0.01	$	&	$	2007	_{-	30	}^{+	25	}$	&	$	10.0	_{-	0.3	}^{+	0.2	}$	&	$	9.90	_{-	0.88	}^{+	3.75	}$	&		Indi,	dist	&	In-situ		\\
69	&	Kepler 7596219	&	$	2.69	\pm	0.02	$&$	19.54	\pm	0.37	$	&	$	7.99	\pm	0.02	$	&	$	0.64	\pm	0.01	$	&	$	3542	_{-	34	}^{+	29	}$	&	$	12.7	_{-	0.2	}^{+	0.2	}$	&	$	9.94	_{-	0.72	}^{+	2.65	}$	&		Indi,	dist	&	In-situ		\\
70	&	Kepler 7908109	&	$	5.84	\pm	0.03	$&$	52.82	\pm	0.53	$	&	$	8.02	\pm	0.01	$	&	$	0.68	\pm	0.01	$	&	$	1709	_{-	14	}^{+	16	}$	&	$	7.6	_{-	0.1	}^{+	0.1	}$	&	$	11.67	_{-	1.15	}^{+	1.94	}$	&		Indi,	no	dist	&	In-situ	\\
71	&	Kepler 7946809	&	$	1.76	\pm	0.02	$&$	11.67	\pm	0.43	$	&	$	8.05	\pm	0.03	$	&	$	0.66	\pm	0.01	$	&	$	3956	_{-	49	}^{+	64	}$	&	$	17.6	_{-	0.4	}^{+	0.4	}$	&	$	9.96	_{-	1.53	}^{+	3.29	}$	&		Seis,	dist	&	In-situ		\\
72	&	Kepler 8350894	&	$	2.00	\pm	0.02	$&$	12.69	\pm	0.29	$	&	$	8.02	\pm	0.02	$	&	$	0.87	\pm	0.01	$	&	$	4142	_{-	92	}^{+	78	}$	&	$	16.6	_{-	0.7	}^{+	0.6	}$	&	$	5.74	_{-	1.53	}^{+	2.84	}$	&		Seis,	no	dist	&	In-situ	\\
73	&	Kepler 8411446	&	$	3.94	\pm	0.10	$&$	28.45	\pm	1.02	$	&	$	7.91	\pm	0.01	$	&	$	0.66	\pm	0.01	$	&	$	3040	_{-	40	}^{+	40	}$	&	$	10.3	_{-	0.2	}^{+	0.3	}$	&	$	10.68	_{-	2.52	}^{+	2.26	}$	&		Seis,	dist	&	In-situ		\\
74	&	Kepler 8544630	&	$	7.55	\pm	0.03	$&$	74.63	\pm	0.51	$	&	$	7.93	\pm	0.01	$	&	$	0.72	\pm	0.01	$	&	$	1742	_{-	23	}^{+	27	}$	&	$	6.4	_{-	0.1	}^{+	0.1	}$	&	$	12.00	_{-	2.33	}^{+	2.01	}$	&		Indi,	no	dist	&	In-situ	\\
75	&	Kepler 9405480	&	$	2.74	\pm	0.01	$&$	22.73	\pm	0.37	$	&	$	8.03	\pm	0.01	$	&	$	0.85	\pm	0.01	$	&	$	1661	_{-	16	}^{+	17	}$	&	$	15.0	_{-	0.3	}^{+	0.2	}$	&	$	3.12	_{-	0.20	}^{+	0.61	}$	&		Seis,	dist	&	In-situ		\\
76	&	Kepler 9407261	&	$	1.57	\pm	0.02	$&$	9.89	\pm	0.48	$	&	$	8.28	\pm	0.04	$	&	$	0.75	\pm	0.01	$	&	$	3703	_{-	67	}^{+	67	}$	&	$	19.7	_{-	0.6	}^{+	0.6	}$	&	$	7.07	_{-	1.86	}^{+	1.97	}$	&		Seis,	dist	&	In-situ		\\
77	&	Kepler 9583607	&	$	3.83	\pm	0.03	$&$	25.25	\pm	0.51	$	&	$	8.02	\pm	0.01	$	&	$	0.67	\pm	0.01	$	&	$	1609	_{-	15	}^{+	16	}$	&	$	10.1	_{-	0.2	}^{+	0.2	}$	&	$	11.63	_{-	2.11	}^{+	1.26	}$	&		Indi,	dist	&	In-situ		\\
78	&	Kepler 9595645	&	$	0.94	\pm	0.01	$&$	5.95	\pm	0.42	$	&	$	8.07	\pm	0.01	$	&	$	0.72	\pm	0.01	$	&	$	1777	_{-	21	}^{+	21	}$	&	$	33.2	_{-	0.7	}^{+	0.6	}$	&	$	1.37	_{-	0.20	}^{+	0.33	}$	&		Seis,	dist	&	In-situ		\\

\end{longtable}

\end{landscape}
\end{onecolumn}



\bsp	
\label{lastpage}
\end{document}